%% file: paper_aa.tex
\documentclass{aa}  

\usepackage{textcomp}
\usepackage{graphicx} % 
\usepackage{amsmath} % 
\usepackage{amssymb} % 
\usepackage{bm} % 
\usepackage{upgreek} %
\usepackage{IEEEtrantools} % 
\usepackage{multirow}
\usepackage[colorlinks=true,allcolors=blue,urlcolor=blue]{hyperref}
\usepackage{txfonts}
\usepackage[normalem]{ulem} 
\usepackage{multirow}
\usepackage[dvipsnames]{xcolor}

\newcommand{\sect}[1]{\text{Sect.~\ref{#1}}}
\newcommand{\fig}[1]{\text{Fig.~\ref{#1}}}

\newcommand{\app}[1]{\text{Appendix~\ref{#1}}}

\newcommand{\abohr}{\mathrm{a_{0}}}
\newcommand{\ratecoeff}{\text{\textlangle}\sigma\varv\text{\textrangle}}
\newcommand{\cfp}{G_{S_{\mathrm{A}},L_{\mathrm{A}}}^{S_{\mathrm{C}},L_{\mathrm{C}}}}
\begin{document} 

\title{Excitation and charge transfer in low-energy
hydrogen atom collisions with neutral carbon and nitrogen
\thanks{Data
available in electronic form at the CDS
via anonymous ftp to 
{\tt cdsarc.u-strasbg.fr (130.79.128.5)}, or 
via~\url{http://cdsarc. u-strasbg.fr/viz-bin/qcat?J/A+A/625/A78} and at
\url{https://github.com/barklem/public-data}.}}
\titlerunning{Low-energy C+H and N+H collisions}
\author{A.~M.~Amarsi\inst{1}
\and
P.~S.~Barklem\inst{2}}
\institute{Max Planck Institute f\"ur Astronomy, K\"onigstuhl 17, 
D-69117 Heidelberg, Germany\\
\email{amarsi@mpia.de}
\and
Theoretical Astrophysics, Department of Physics and Astronomy, 
Uppsala University, Box 516, SE-751 20 Uppsala, Sweden\\
\email{paul.barklem@physics.uu.se}}

\abstract{Low-energy inelastic collisions with neutral hydrogen atoms
are important processes in stellar atmospheres,
and a persistent source of uncertainty in non-LTE modelling
of stellar spectra. We have calculated and studied excitation and charge transfer
of \ion{C}{I} and of \ion{N}{I}~due to such collisions.
We used a previously presented method
that is based on an asymptotic two-electron linear combination of atomic
orbitals (LCAO) model of ionic-covalent interactions for the adiabatic
potential energies, combined with the multichannel Landau-Zener model
for the collision dynamics. 
We find that charge transfer processes typically lead to 
much larger rate coefficients than excitation processes do,
consistent with studies of other atomic species.
Two-electron processes were considered
and lead to non-zero rate coefficients
that can potentially impact
statistical equilibrium calculations.
However, they were included in the model
in an approximate way, via an estimate for the 
two-electron coupling that was presented earlier in the literature:
the validity of these data should be checked in a future work.}


\maketitle
%-------------------------------------------------------------------------------
\section{Introduction}
\label{introduction}

Carbon and nitrogen, 
respectively the second and fourth most abundant metals in the 
Sun, are among the most important elements in 
modern astrophysics.
They play a key role in stellar physics as catalysts in the CNO-cycle.
As a result of this cycle their abundances are interlinked,
and their abundances in the surfaces of stars that have
dredged-up or mixed C-poor, N-rich material from the interior
has been used to study stellar evolution 
\citep[e.g.][]{2000A&A...354..169G,2005A&amp;A...430..655S}.
Since the amount of mixing is sensitive to stellar mass,
[C/N] measurements are also good indicators of the ages of late-type giants
\citep[e.g.][]{2015MNRAS.453.1855M,2015A&A...583A..87S,
2016MNRAS.456.3655M}.
Finally, C-N abundance variations in globular cluster stars
of the same evolutionary stages shed light on
their formation histories
and different populations \citep[e.g.][]{2012A&ARv..20...50G}.

As such, it is highly desirable to determine
carbon and nitrogen abundances as accurately as possible.
Lines of \ion{C}{I} and \ion{N}{I} are commonly used
to measure carbon and nitrogen abundances,
in the Sun \citep[e.g.][]{2009ARA&amp;A..47..481A,
2009A&A...498..877C,2010A&amp;A...514A..92C}
and early- and late-type stars
\citep[e.g.][]{1978ApJ...223..937T,1981ApJ...250..262C,
1994PASJ...46...53T,
2001A&amp;A...379..936P,2001A&A...379..955P,
2002A&A...381..982S,
2005PASJ...57...65T,2011MNRAS.410.1774L,
2015MNRAS.446.3447L}.

It is well-known that 
the formation of \ion{C}{I} and \ion{N}{I} lines
in stellar atmospheres is
sensitive to departures from local thermodynamic equilibrium (LTE).
Non-LTE modelling requires a complete description of all the relevant
radiative and collisional bound-bound and bound-free transitions,
and historically the lack of reliable data for inelastic collisions
with electrons and with hydrogen
has been a dominant source of uncertainty.
The situation has recently improved for 
\ion{C}{I} and \ion{N}{I}, for which
recent $B$-spline $R$-matrix (BSR) calculations
have put the rates of electron-impact excitation and ionisation
on firmer footing \citep{2013PhRvA..87a2704W,2014PhRvA..89f2714W}.

Concerning inelastic collisions with neutral hydrogen,
full quantum-scattering calculations within
the Born-Oppenheimer approach for \ion{Li}{I},
\ion{Na}{I}, and \ion{Mg}{I} have
indicated the importance of excitation and charge transfer through
avoided ionic crossings \citep{1999PhRvA..60.2151B,
2010PhRvA..81c2706B,2012PhRvA..85c2704B,
2003PhRvA..68f2703B,2011JPhB...44c5202G}.
As discussed in previous papers in this series
\citep{2018A&A...612A..90B,2018A&A...610A..57B},
such calculations are time-consuming and complicated. As a result, several asymptotic methods have been developed to model this mechanism:
namely, the methods of \citet{2013PhRvA..88e2704B} 
based on semi-empirical couplings \citep{1971ApOpt..10.1848O};
and of \citet{2016PhRvA..93d2705B}, with theoretical couplings derived from
a theoretical two-electron linear combination of atomic orbitals
\citep[LCAO;][]{1974MolPh..27..159G,1977MolPh..33..793A}.
These asymptotic methods reproduce 
the full quantum scattering results reasonably well,
at least for 
for the species \ion{Li}{I}, \ion{Na}{I}, and \ion{Mg}{I}
and for processes involving states of low excitation
potential having large rate coefficients
(that tend to be of importance
for statistical equilibrium calculations involving these species).

Here, we present calculations for inelastic collisions with neutral
hydrogen using the LCAO model of \citet{2016PhRvA..93d2705B},
for \ion{C}{I} and \ion{N}{I}.
We present the method in \sect{method},
and discuss the results of the calculations in 
\sect{results}.
We summarise these data and remark on their 
applicability in \sect{conclusion}.

%-------------------------------------------------------------------------------
\section{Method}
\label{method}

\subsection{Input data and assumptions}
\label{method_lcao}

Calculations of the transitions through avoided ionic crossings
(via the $\mathrm{C^{+}+H^{-}}$~and 
$\mathrm{N^{+}+H^{-}}$~ionic configurations)
were performed following the method
described in \citet{2016A&amp;ARv..24....9B},
where it is applied to
\ion{Li}{I}, \ion{Na}{I}, \ion{Mg}{I},
and \ion{Ca}{I}. The method was subsequently applied to
\ion{O}{I} \citep{2018A&A...610A..57B},
\ion{Fe}{I} \citep{2018A&A...612A..90B},
as well as \ion{K}{I} and \ion{Rb}{I} \citep{2018MNRAS.473.3810Y}.
As in those works, the production runs were based on 
adiabatic potential energies and 
coupling parameters derived from the LCAO model.
The potential energies were calculated for internuclear
distances $3.0 \leq R / \abohr \leq 400.0$: at shorter 
internuclear distances,
other couplings not considered here (rotational 
and spin-orbit couplings) are likely to be important, while 
avoided ionic crossings at larger internuclear distances
do not give rise to large rate coefficients.
The collisional cross-sections were 
calculated using the multichannel Landau-Zener model,
for collision energies from threshold up to $100\,\mathrm{eV}$,
from which rate coefficients were calculated 
for temperatures $1000 \leq T / \mathrm{K} \leq 20000$.

We present the input data and the considered symmetries in
\app{appendix_lcao}.  To summarise,
the energies for these low-lying levels were taken
from \citet{gallagher1993tables} via
the NIST Atomic Spectra Database \citep{NIST_ASD}.
Fine structure was collapsed and $g$-weighted energies were
adopted. Ionic states involving
$\mathrm{C^{+}\,2p\,^{2}P^{o}}$,
and $\mathrm{N^{+}\,2p^{2}\,^{3}P}$~and
$\mathrm{N^{+}\,2p^{2}\,^{1}D}$, were included
for carbon and nitrogen respectively.  Initially
we had considered a greater number of ionic states,
however the corresponding transitions have avoided
ionic crossings at very short internuclear distances,
and have a negligible impact on the current model.
As a consequence of omitting some ionic states,
the coefficients of fractional parentage here do
not always satisfy the 
normalisation condition $\sum[(\cfp)^2] = 1$, 
summing over cores $\mathrm{C}$.

Following \citet{2018A&A...610A..57B},
two-electron processes were included in an approximate way,
by calculating the analogous one-electron
couplings and applying a scaling factor to it
based on the internuclear distance, 
following Eq.~1 of \citet{2017A&A...606A.106B}
(see also earlier work by \citealt{1993PhRvA..48.4299B}
and \citealt{2016A&A...593A..27Y}).
Two-electron processes involving hydrogen in the first excited state
(energy of $10.20\,\mathrm{eV}$) were
considered potentially important, given
the high ionisation limits of 
\ion{C}{I} (ionisation limit of $11.26\,\mathrm{eV}$)
and \ion{N}{I} (ionisation limit of $14.53\,\mathrm{eV}$).
They were considered for the 
$\mathrm{2s^{2}.2p^{2}\,{^3}P}$~ground state of \ion{C}{I}, and for the 
$\mathrm{2s^{2}.2p^{3}\,{^4}S^{o}}$~ground state
and $2\mathrm{s^{2}.2p^{3}\,{^2}D^{o}}$~(energy of $2.38\,\mathrm{eV}$) 
first excited state of \ion{N}{I}.
These proceed via interaction of ionic states 
$\mathrm{X^{+}\,2s^{2}.2p^{a} + H^{-}}$, with 
covalent states $\mathrm{2s^{2}.2p^{a+1} + H\left(n=2\right)}$.
In the current model these are interpreted as two-electron processes,
wherein the two electrons initially located on $\mathrm{H^{-}}$ transfer onto
$\mathrm{X^{+}}$, and then,
due to the lack of an accepting state, one transfers back,
resulting in $\mathrm{H\left(n=2\right)}$.

Two-electron processes with hydrogen in its ground state
were also considered,
for the $\mathrm{2s.2p^{3}}$~configurations of \ion{C}{I},
and for the $\mathrm{2s.2p^{4}}$~configuration of \ion{N}{I}.
These proceed via interaction of ionic states 
$\mathrm{X^{+}\,2s^{2}.2p^{a}+H^{-}}$, with covalent states 
$\mathrm{2s.2p^{a+2} + H\left(n=1\right)}$\footnote{Hereafter,
it is implied that $\mathrm{H}$~is in its ground
state, unless otherwise indicated.}.
These processes present added complications.
The assumption implicit in the model, that the core is in 
a frozen configuration, breaks down.
As a result, the angular momentum coupling term, $C$~in 
Eq.~19 and Eq.~20 of \citet{2016PhRvA..93d2705B},
usually calculated for one particular frozen core,
is here undefined, as is the electron binding energy
and the coefficient of fractional parentage.
To be able to obtain some estimate with the existing model,
a number of approximations were made for the covalent states:
namely, the angular momentum coupling terms were set to unity,
the electron binding energies were chosen to correspond to the core 
for which the coefficient of fractional parentage was largest,
and the coefficients of fractional parentage were set to unity.
The resulting rate coefficients may be
considered indicative of the potential for such processes to be
of significance.

\subsection{Alternative ionic channels}
\label{method_alt}

Only transitions via the $\mathrm{C^{+}+H^{-}}$~(asymptotic limit
at $10.6\,\mathrm{eV}$~relative to the neutral ground states) and
$\mathrm{N^{+}+H^{-}}$~($13.8\,\mathrm{eV}$) ionic configurations were
considered here.  Other ionic configurations and channels are 
also possible. In particular, since 
\ion{C}{I} and \ion{N}{I} have large 
ionisation potentials, comparable to 
that of \ion{H}{I},
transitions could occur via the $\mathrm{C^{-}+H^{+}}$~($12.3\,\mathrm{eV}$)
and $\mathrm{N^{-}+H^{+}}$~($13.7\,\mathrm{eV}$) ionic configurations.
We note that
unless the carbon or nitrogen atom in the covalent state
is in the ground configuration
($2p^2$ or $2p^3$, respectively),
this interaction must involve two active electrons.
For example, if the electron on 
the carbon atom is in the $3s$~orbital, then the process 
would involve both the transition of this electron from the $3s$~orbital
to the $2p$~orbital,
and in addition the transfer of the electron from the 
hydrogen atom to the $2p$~carbon core,
so as to form C$^-$ ($2p^3$).
The approximate scaling factor mentioned in
\sect{method_lcao} is probably not appropriate in this case,
because the electron is here transferring to the carbon atom
in the covalent state, rather than from it, and the carbon atom 
undergoes rearrangement of the core electrons.

The possible importance of these alternative
ionic configurations and channels was    tested for carbon.
Mutual neutralisation cross-sections
into the ground configurations from
$\mathrm{C^{+}+H^{-}}$ and $\mathrm{C^{-}+H^{+}}$~were
estimated using the two channel Landau Zener model
together with semi-empirical couplings from 
\citet{1971ApOpt..10.1848O}, for energies in the range 
$0.1$--$1.0\,\mathrm{eV}$.
The cross-sections from $\mathrm{C^{+}+H^{-}}$ are two
orders of magnitude larger than from $\mathrm{C^{-}+H^{+}}$ at 
$1.0\,\mathrm{eV}$, reaching four
orders of magnitude at $0.1\,\mathrm{eV}$.  
This is understandable from the fact that the
crossings for $\mathrm{C^{-}+H^{+}}$ occur at shorter internuclear distances
than for $\mathrm{C^{+}+H^{-}}$, as well as the 
binding energies of $\mathrm{C^{-}}$~and
$\mathrm{H}$~(the initial and final states of the
active electron in the former case)
being larger than for $\mathrm{H^{-}}$ and $\mathrm{C}$~(the 
initial and final states of the
active electron in the latter case).
Thus, this estimate, combined with the
expectation that the cases involving two-electron processes
will generally be
less efficient than the cases involving 
one-electron processes, suggests that
$\mathrm{C^{+}+H^{-}}$ is unlikely to be important compared to
$\mathrm{C^{-}+H^{+}}$ at collision energies of interest.

The case of nitrogen is
complicated by the fact that 
$\mathrm{N^{-}}$ has a negative electron affinity 
($-0.07\,\mathrm{eV}$),
and thus the assignment of the binding energy of the active electron --- required to
derive the semi-empirical coupling --- is not obvious. 
The ionic crossings for $\mathrm{N^{-}+H^{+}}$~take place
at slightly larger internuclear distance than for $\mathrm{N^{+}+H^{-}}$.
However, the expectation that the cases involving two-electron processes 
are generally less efficient than the cases involving one-electron processes
still applies.

In summary, the
$\mathrm{C^{-}+H^{+}}$ and $\mathrm{N^{-}+H^{+}}$ configurations 
were neglected in these calculations.
These alternative channels are expected to be of lesser importance,
however their possible influence perhaps makes these
calculations somewhat more uncertain,
compared to calculations for atoms with small ionisation
potentials 
relative to that of \ion{H}{I}.  
Quantum chemistry calculations would be useful to
investigate the influence of these ionic configurations.

\begin{figure*}
    \begin{center}
        \includegraphics[scale=0.31]{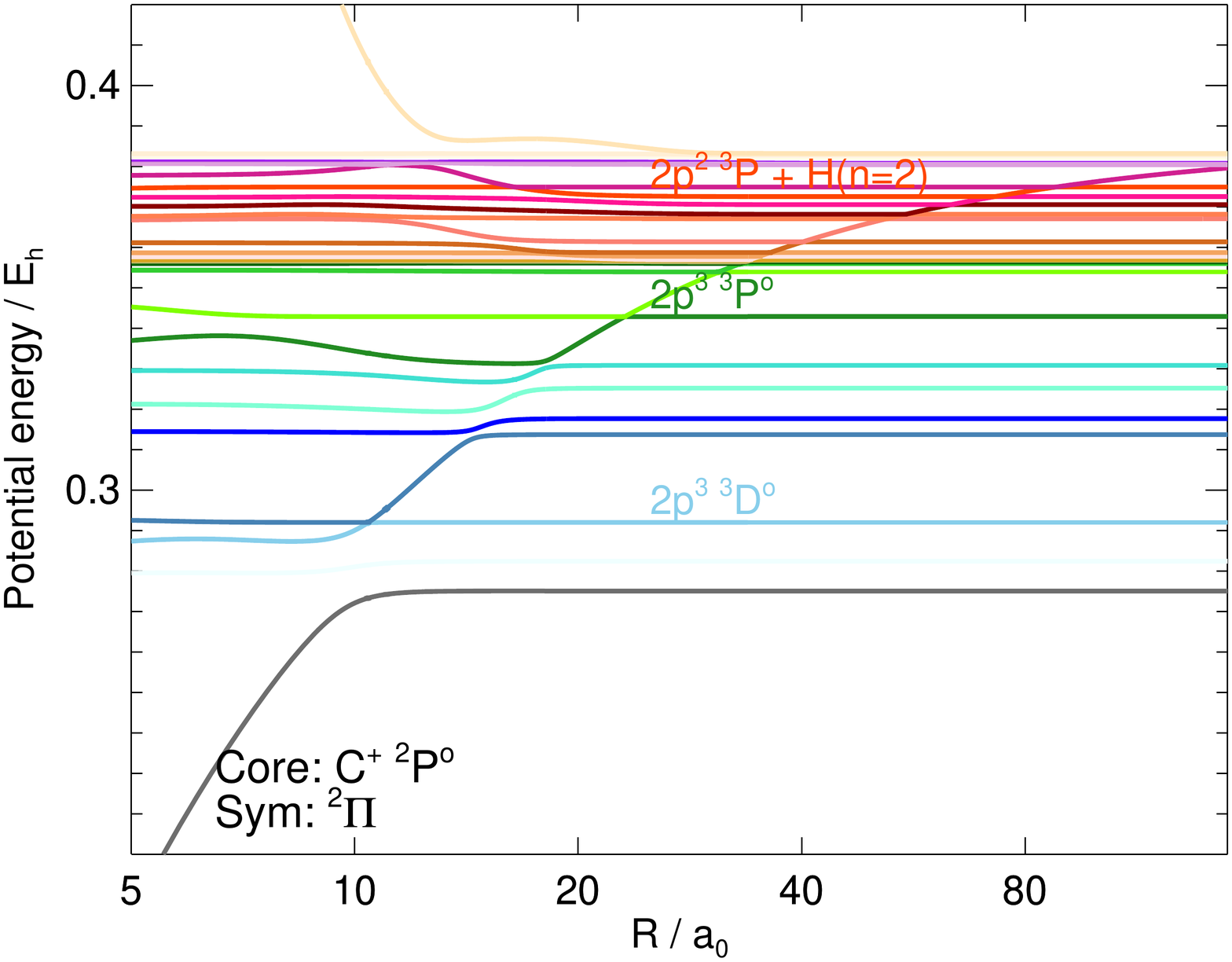}
        \includegraphics[scale=0.31]{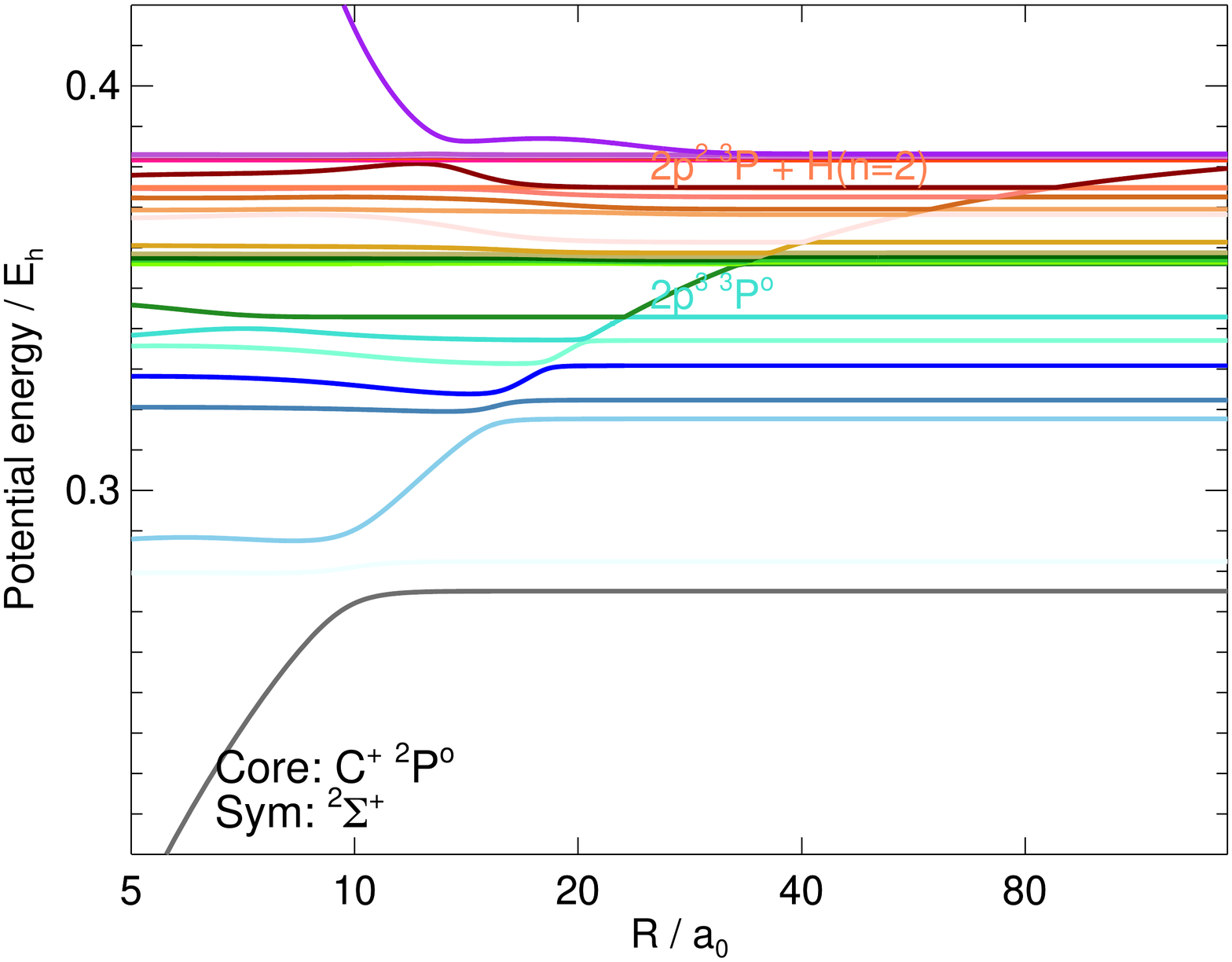}
        \includegraphics[scale=0.31]{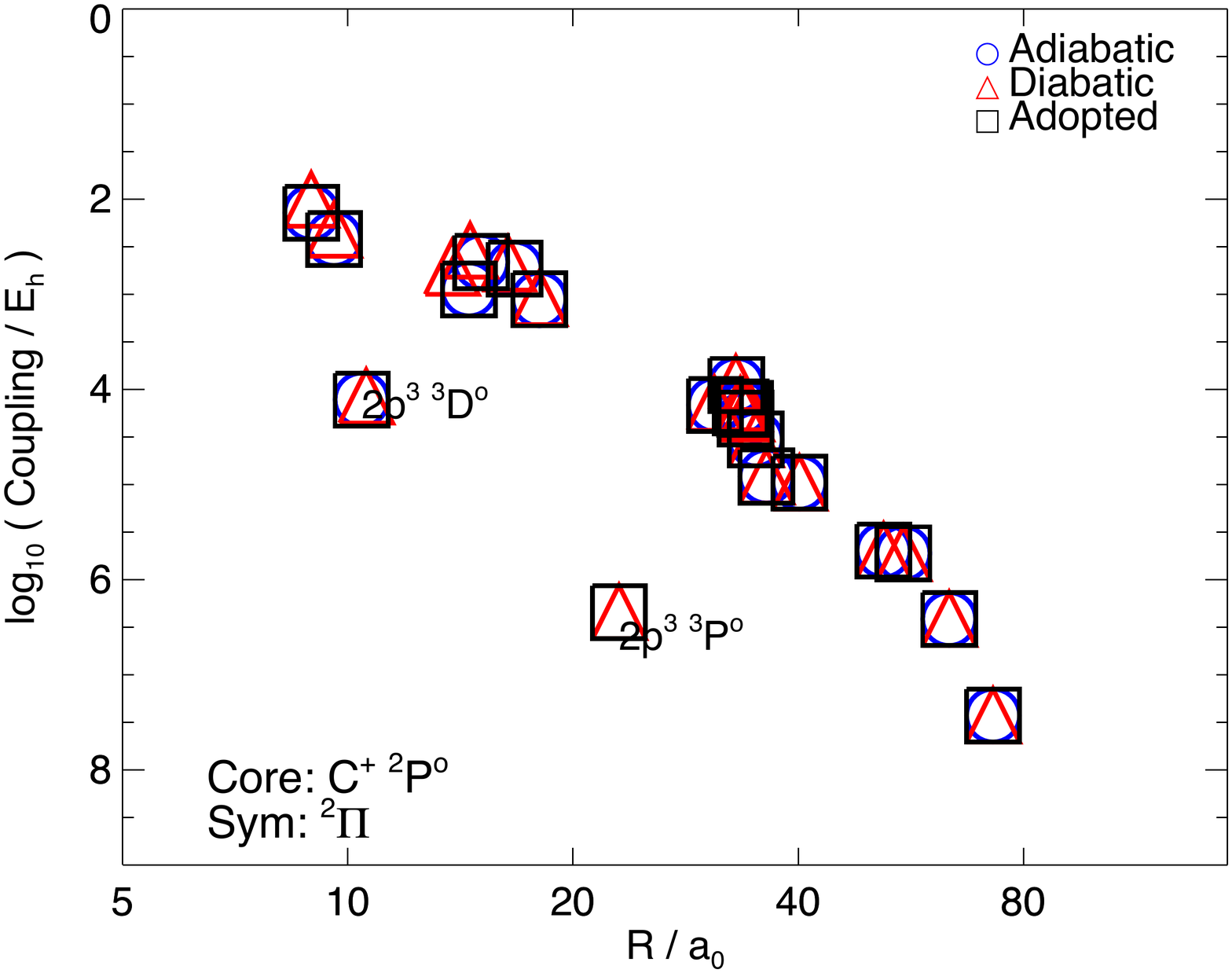}
        \includegraphics[scale=0.31]{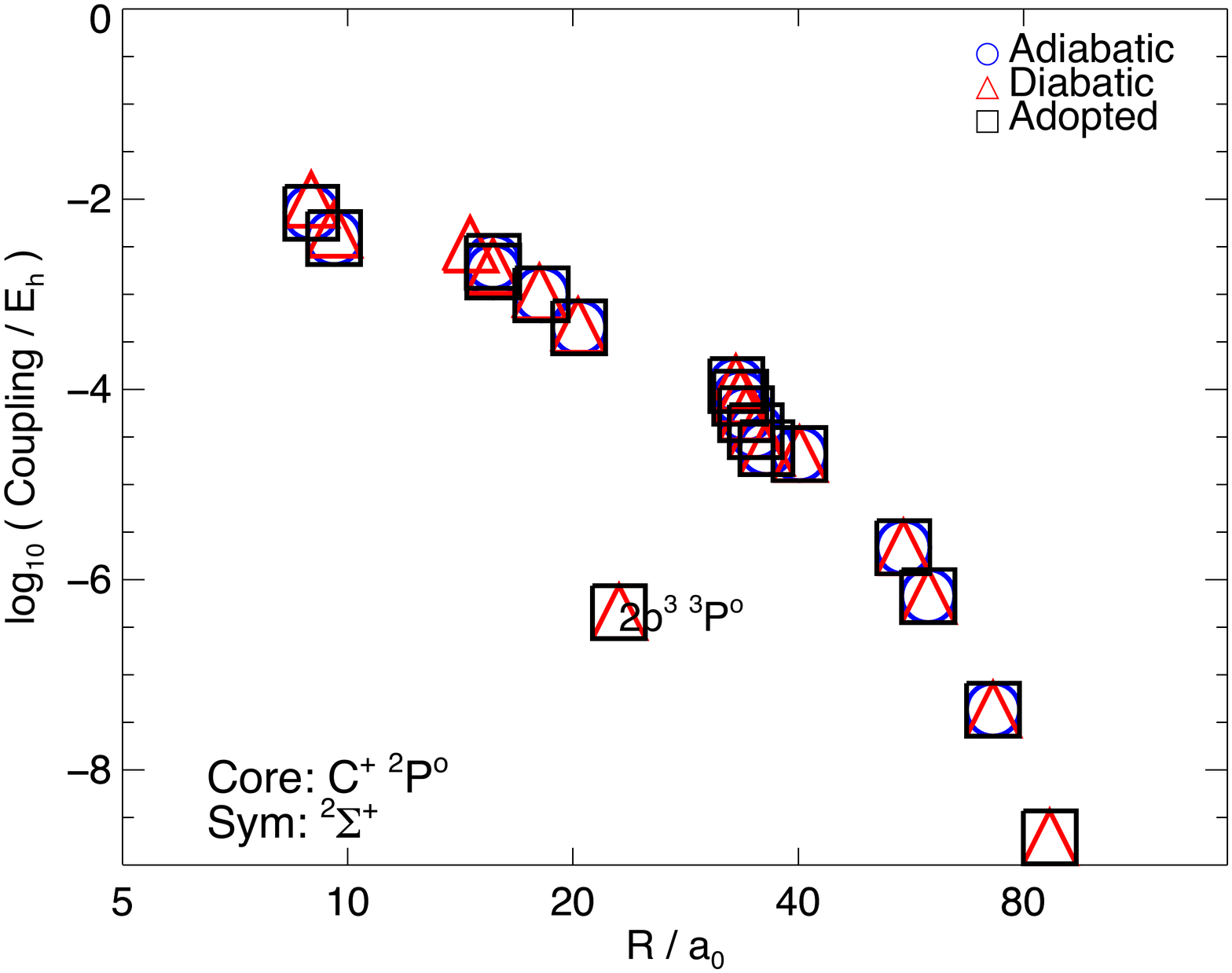}
        \caption{Adiabatic potential energies as a 
        function of internuclear distance
        (top), and corresponding couplings at 
        the crossing radius (bottom),
        in atomic units 
        (Hartree energy $E_{\mathrm{h}}$~and
        Bohr radius $\abohr$, respectively),
        for symmetries involving the ground state ionic
        cores of carbon.
        States included in the model via two-electron processes
        have been labelled. The lowest-lying states
        have been truncated from the upper plots.
        For the \ion{C}{I} $\mathrm{2s.2p^{3}\,^{3}P^{o}}$~state,
        the coupling is very small; the diabatic value was
        judged to be more reliable and the adiabatic value is not shown.}
        \label{fig:lcao_pc_carbon}
    \end{center}
\end{figure*}

\begin{figure*}
    \begin{center}
        \includegraphics[scale=0.31]{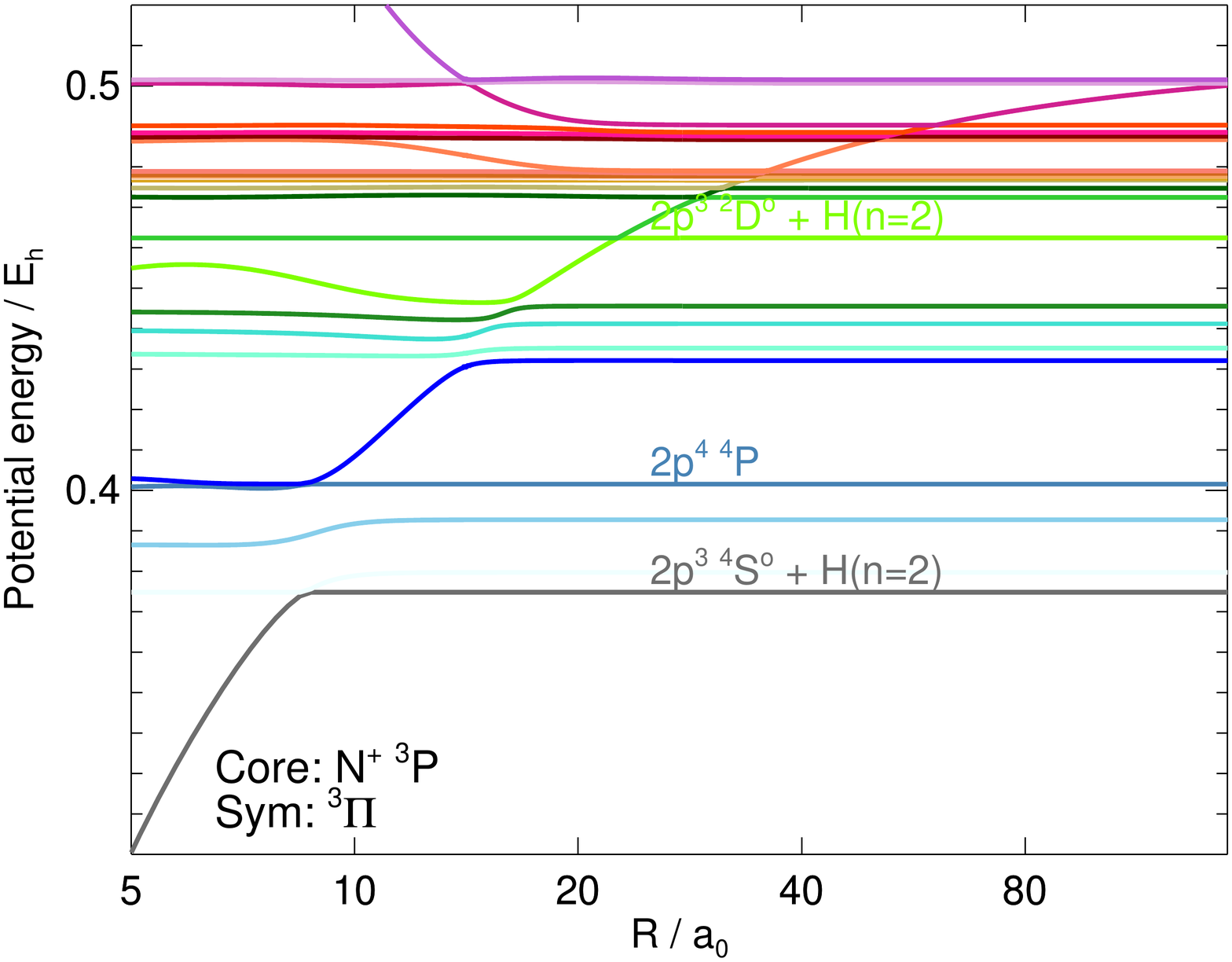}
        \includegraphics[scale=0.31]{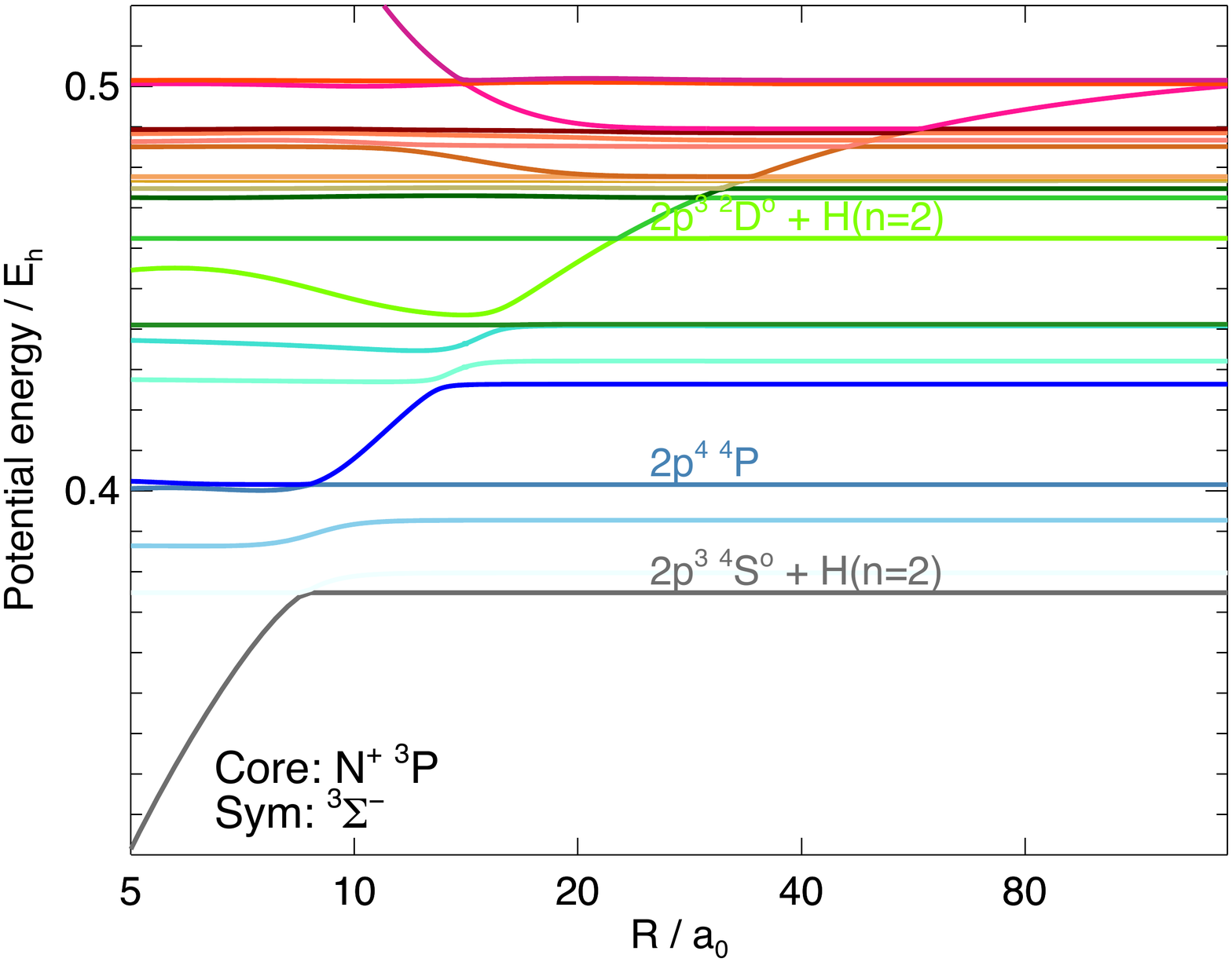}
        \includegraphics[scale=0.31]{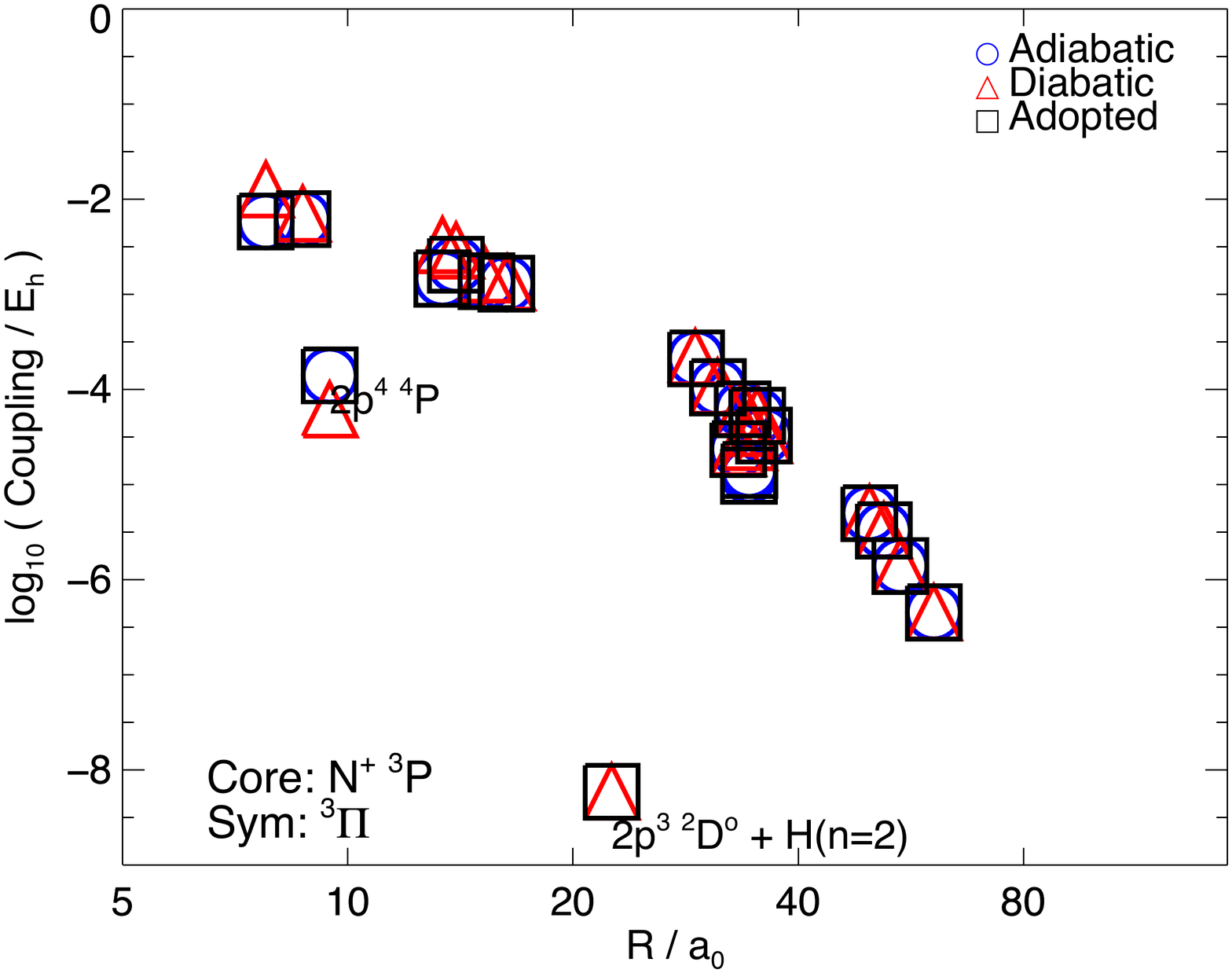}
        \includegraphics[scale=0.31]{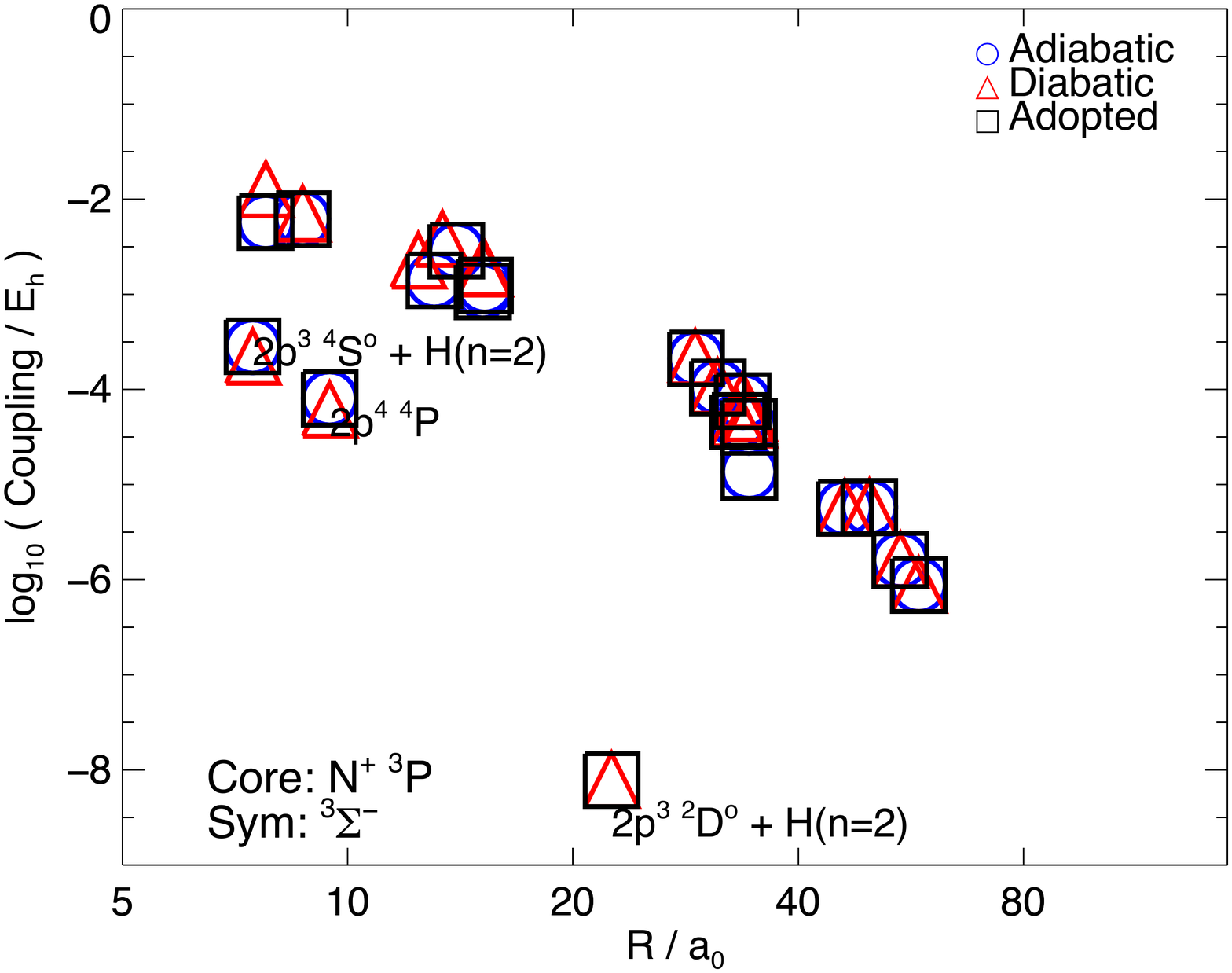}
        \caption{Adiabatic potential energies as 
        a function of internuclear distance
        (top), and corresponding couplings at 
        the crossing radius (bottom), in atomic units
        (Hartree energy $E_{\mathrm{h}}$~and
        Bohr radius $\abohr$, respectively),
        for symmetries involving the ground state ionic
        cores of nitrogen.
        States included in the model via two-electron processes
        have been labelled. The lowest-lying states
        have been truncated from the upper plots.
        For the \ion{N}{I}
        $\mathrm{2s^{2}.2p^{3}\,^{2}D^{o}+H\left(n=2\right)}$~state
        the coupling is very small; the diabatic value was
        judged to be more reliable and the adiabatic value is not shown.}
        \label{fig:lcao_pc_nitrogen}
    \end{center}
\end{figure*}

\begin{figure*}
    \begin{center}
        \includegraphics[scale=0.66]{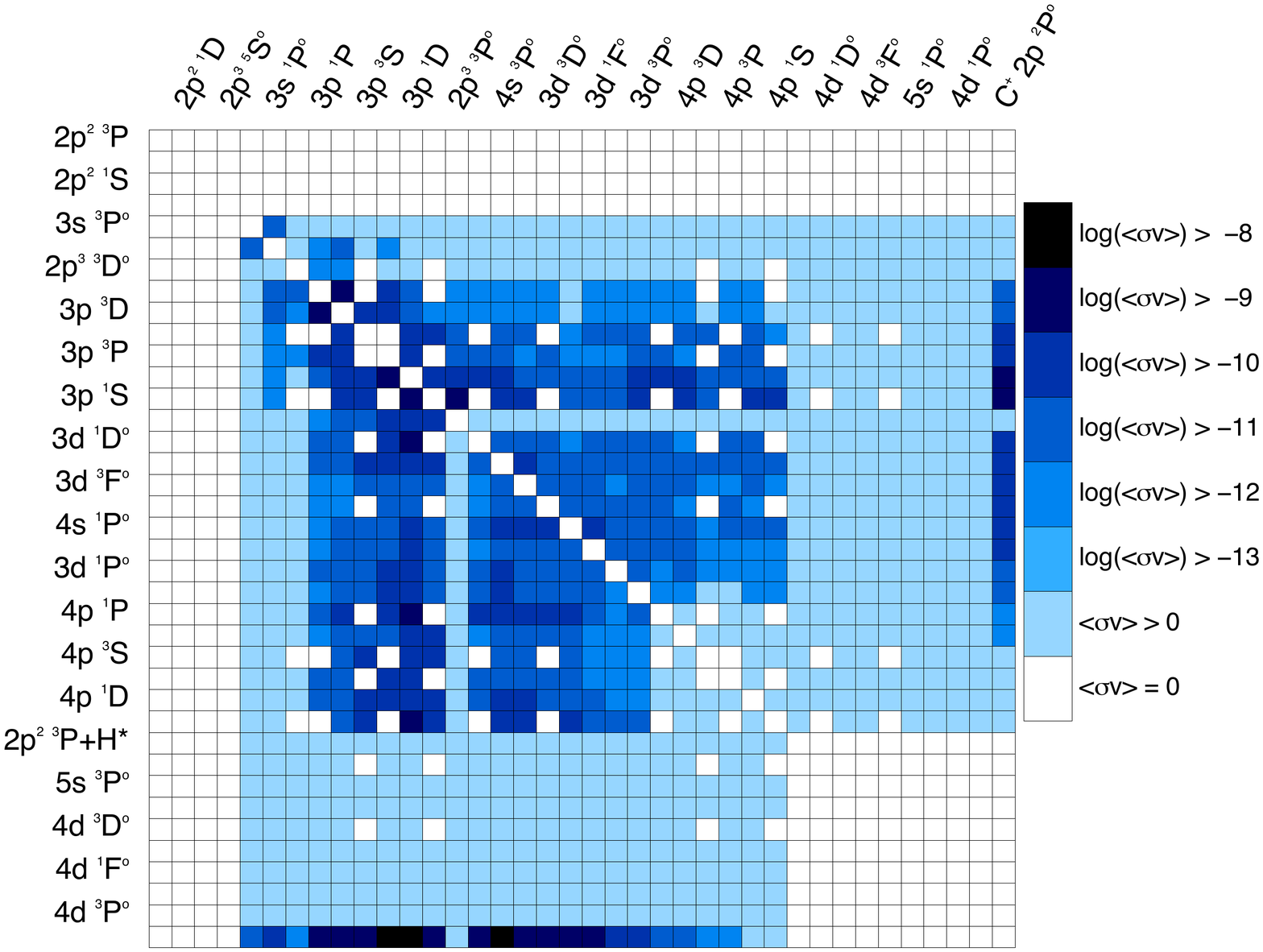}
        \caption{Graphical representation of the rate coefficient matrices
        $\ratecoeff$~at $6000\,\mathrm{K}$~for C+H.
        Rows give initial states, columns give final states,
        in order of increasing covalent energies,
        with every other state being labelled on each axis.
        The rate coefficients are in $\mathrm{cm^{3}\,s^{-1}}$.}
        \label{fig:lcao_grid_c}
    \end{center}
\end{figure*}

\begin{figure*}
    \begin{center}
        \includegraphics[scale=0.66]{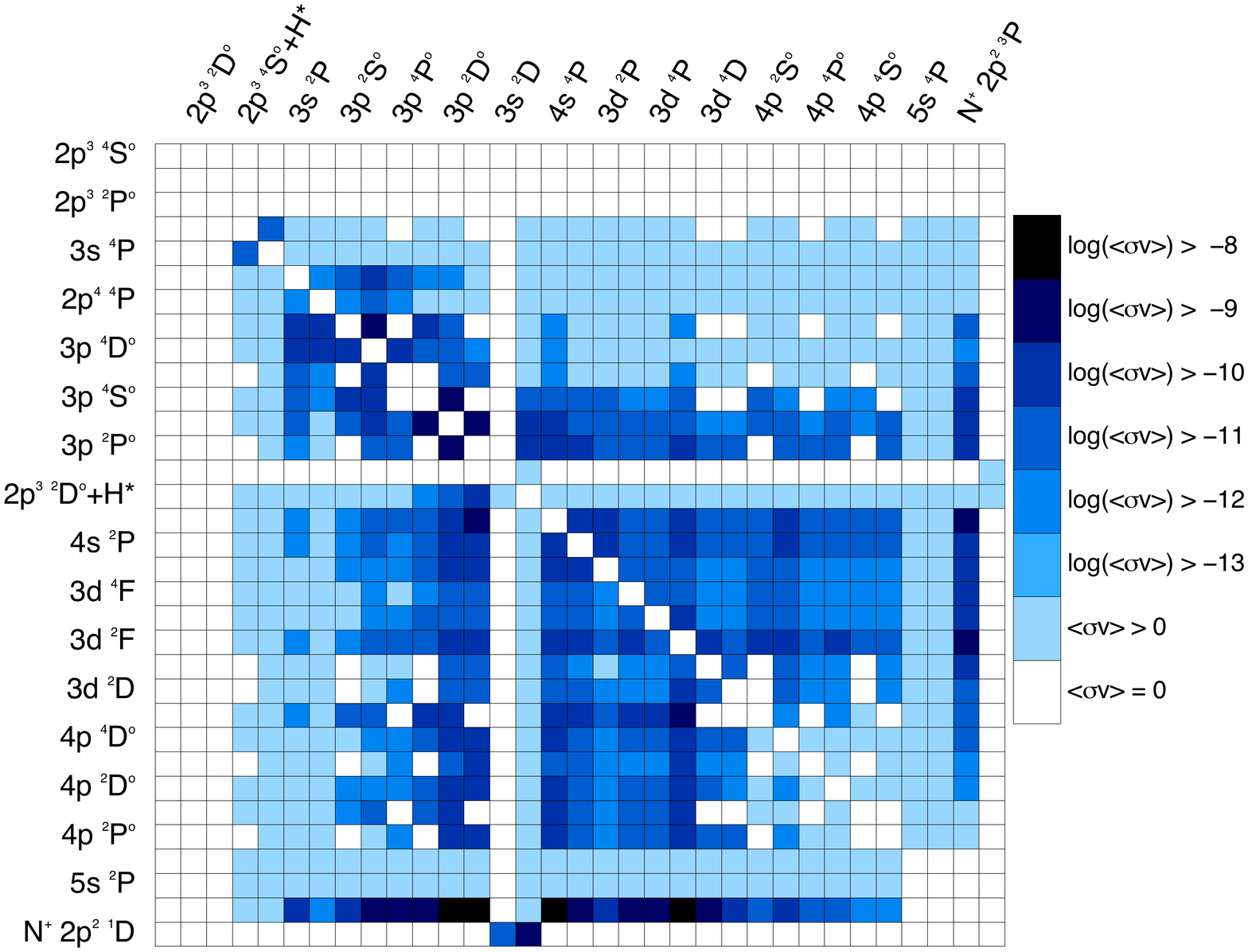}
        \caption{Graphical representation of the rate coefficient matrices
        $\ratecoeff$~at $6000\,\mathrm{K}$~for N+H.
        Rows give initial states, columns give final states,
        in order of increasing covalent energies,
        with every other state being labelled on each axis.
        The rate coefficients are in $\mathrm{cm^{3}\,s^{-1}}$.}
        \label{fig:lcao_grid_n}
    \end{center}
\end{figure*}

\begin{figure}
    \begin{center}
        \includegraphics[scale=0.31]{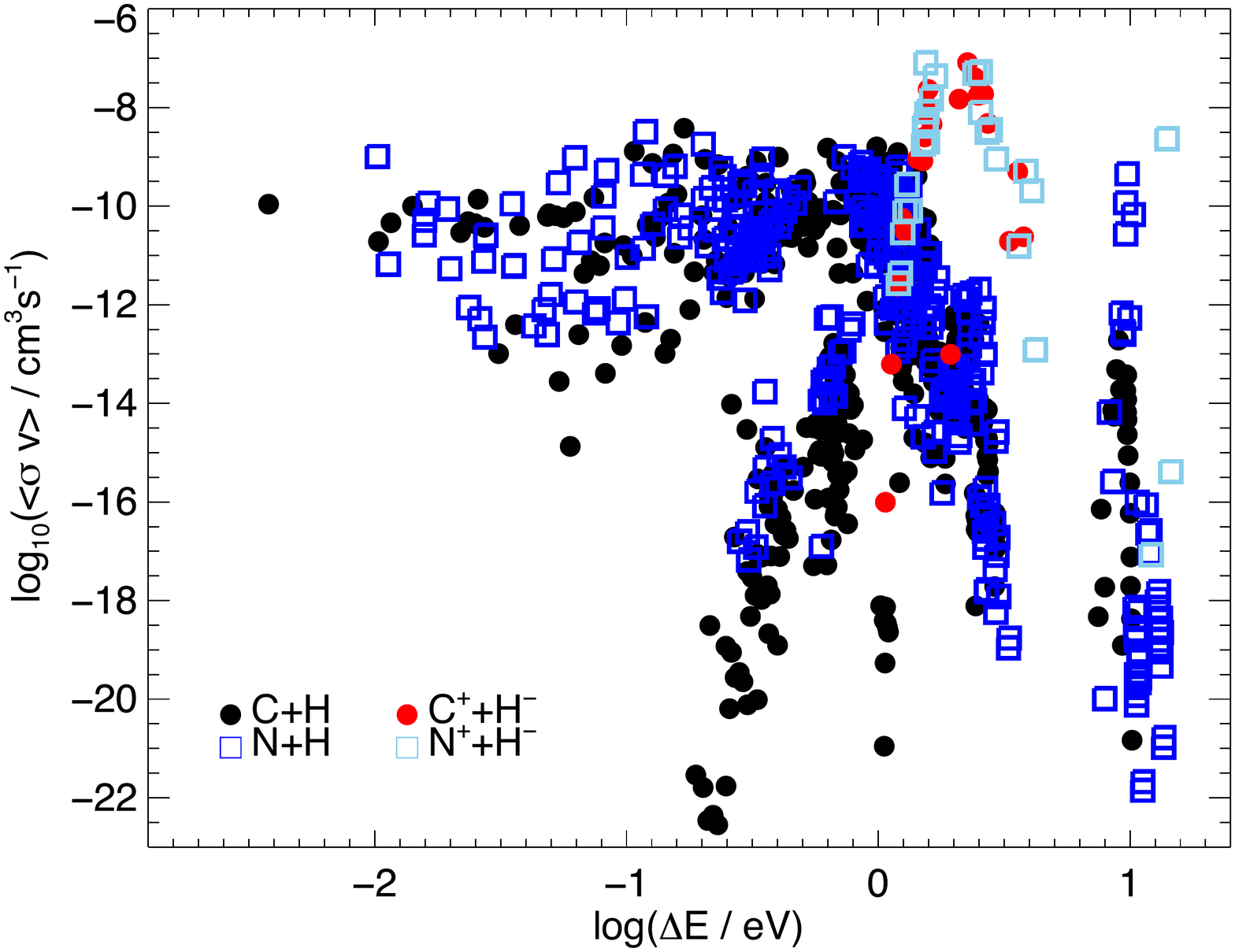}
        \caption{De-excitation and mutual neutralisation
        rate coefficients $\ratecoeff$~at $6000\,\mathrm{K}$~for 
        C+H and N+H.}
        \label{fig:rates_de}
    \end{center}
\end{figure}

%-------------------------------------------------------------------------------
\section{Results}
\label{results}

\subsection{Potentials and couplings}
\label{results_lcao_a}

In the top rows of Figs.~\ref{fig:lcao_pc_carbon} and 
\ref{fig:lcao_pc_nitrogen}
we illustrate the adiabatic potential energies
as a function of internuclear distance, for covalent states 
involving the ground state ionic cores of \ion{C}{II} and 
\ion{N}{II}.
First, we note that 
the covalent states involving ground-state hydrogen
and the four lowest-lying levels
of \ion{C}{I} ($\mathrm{2s^{2}.2p^{2}\,^{3}P\text{/}
^{1}D\text{/}^{1}S\text{, and } 2s.2p^{3}\,^{5}S^{o}}$),
and the three lowest-lying levels of \ion{N}{I}
($\mathrm{2s^{2}.2p^{3}\,^{4}S^{o}\text{,}
\,^{2}D^{o}\text{,}\,^{2}P^{o}}$),
do not have any avoided ionic crossings in the current model,
for $R \geq 3\,\abohr$. Their adiabatic potential energies
are not shown in the figures, lying below the horizontal axes.
The plots show the covalent states involving
ground-state hydrogen, and levels of 
intermediate and high excitation potential
of \ion{C}{I} and \ion{N}{I}, namely
\ion{C}{I} $\mathrm{2s^{2}.2p.3s\,^{3}P^{o}}$~and
\ion{N}{I} $\mathrm{2s^{2}.2p{^2}.3s\,^{4}P}$, and
higher.  These have avoided ionic crossings 
at intermediate internuclear distances.

The top rows of 
Figs.~\ref{fig:lcao_pc_carbon} and \ref{fig:lcao_pc_nitrogen}
also show the adiabatic potential energies of the 
covalent states involving excited hydrogen
and the lowest energy state of \ion{C}{I},
as well as excited hydrogen
and the two lowest energy states of \ion{N}{I}.
The former (\ion{C}{I}) 
is relatively highly excited, and as such has its
avoided ionic crossing at larger internuclear distance
(close to $80\,\abohr$). 
The latter two (\ion{N}{I})
have avoided ionic crossings at short and intermediate internuclear distances
(around $7\,\abohr$~and $20\,\abohr$, respectively).

In the bottom rows of 
Figs.~\ref{fig:lcao_pc_carbon} and \ref{fig:lcao_pc_nitrogen} 
we illustrate the corresponding ionic-covalent coupling coefficients,
calculated using both adiabatic and diabatic potential energies
in the Landau-Zener formalism. As discussed in
Sect.~II.B.3 of \citet{2016PhRvA..93d2705B}, the adiabatic
values are generally expected to be most reliable. However,
when the couplings are very small, they are difficult to
accurately determine from the potentials in the adiabatic representation, and
the minimum coupling that can be resolved 
at a given internuclear distance is limited by the numerical
properties of the calculations.  In earlier work, this has
only occurred for crossings at large internuclear distances
and so was dealt with by adopting diabatic values at $R>50\,\abohr$.
In this work, small couplings also occur at intermediate internuclear
distances, for two-electron processes (for example, for the
\ion{N}{I} $\mathrm{2s^{2}.2p^{3}\,^{2}D^{o}}$~state
in \fig{fig:lcao_pc_nitrogen}). 
In cases in which a reliable adiabatic value
cannot be obtained, the diabatic values were adopted.

%-------------------------------------------------------------------------------
%-------------------------------------------------------------------------------

\subsection{Rate coefficients}
\label{results_lcao_b}

In Figs.~\ref{fig:lcao_grid_c}
and \ref{fig:lcao_grid_n} we illustrate the resulting total
rate coefficients at $6000\,\mathrm{K}$.
The rates are in matrix form: the lower diagonal 
represents downwards processes 
(de-excitation and mutual neutralisation),
and the upper diagonal represents
upwards processes (excitation and ion pair production). 
In \fig{fig:rates_de} we plot the downwards
rate coefficients as functions of the logarithm of the
transition energy $\Delta E$.

The rate coefficients are correlated with the excitation 
potential. The largest rates tend to involve
states of intermediate and moderately-high excitation potential:
for excitation and de-excitation, 
this corresponds to the squares in the middle of
Figs.~\ref{fig:lcao_grid_c} and \ref{fig:lcao_grid_n},
while for ion pair production and mutual neutralisation, 
this corresponds to the squares in the middle of
the last rows and columns of these figures.
The rate coefficients for transitions involving the 
states of lowest excitation potential 
(shown on the first 
four rows and columns of 
\fig{fig:lcao_grid_c} and 
the first three rows and columns of
\fig{fig:lcao_grid_n}) are zero.  As discussed in
\sect{results_lcao_a}, the 
corresponding avoided ionic crossings
happen at very short internuclear distances, beyond the regime of
the present model and consequently do not give rise to any couplings.
On the other hand, for the states of highest excitation potential,
the rate coefficients for 
excitation as well as charge transfer
are generally small ($\ratecoeff<10^{-13}\,\mathrm{cm^{3}\,s^{-1}}$),
if not zero.  The corresponding avoided ionic crossings
happen at large internuclear distances, giving rise to smaller
couplings:  in the quantum tunnelling interpretation,
this can be interpreted as a result of the 
electron tunnelling probability 
dropping off with increasing 
width of the potential barrier.

The rates also have a strong dependence on the transition
energy. The mutual neutralisation rates
in \fig{fig:rates_de} have a sharp peak at around 
$\log({\Delta E / \mathrm{eV}})\approx0.3$,
or transition energies of around $2\,\mathrm{eV}$;
the de-excitation rates have a broader peak, at around 
$\log({\Delta E / \mathrm{eV}})\approx0$,
or transition energies of around $1\,\mathrm{eV}$.
In the case of de-excitation, some structure in the rate
coefficients is clearly visible, 
generally reflecting their dependence also on the excitation
potential as discussed above.
This behaviour --- the rate coefficients peaking
at particular energies --- has been previously noted
in the context of other species \citep[e.g.][]{2018A&A...612A..90B},
and is exploited for calculating more simplified models
for inelastic hydrogen collisions
\citep[e.g.][]{2017A&A...606A.147B,2018A&A...618A.141E}.

Charge transfer processes
typically give the largest rates,
as has been found in previous studies of other species.
This can be seen from the dark squares in
the last rows and columns of 
\fig{fig:lcao_grid_c} and \fig{fig:lcao_grid_n},
and from the spike in mutual neutralisation rate coefficients
at $\log({\Delta E / \mathrm{eV}})\approx0.3$~in
\fig{fig:rates_de}.
This result can be understood by noting that
in the current model,
charge transfer 
corresponds to a single event, of an electron moving 
from the carbon or nitrogen atom, onto the hydrogen atom
(or vice versa).  In contrast, excitation requires
a second event: that electron moving back onto 
the carbon or nitrogen atom in some different energy state.

The covalent states involving excited hydrogen,
namely \ion{C}{I} $\mathrm{2s^{2}.2p^{2}\,^{3}P + H\left(n=2\right)}$,
and \ion{N}{I} 
$\mathrm{2s^{2}.2p^{3}\,^{4}S^{o}\text{/}^{2}D^{o} 
+ H\left(n=2\right)}$,
correspond to the twenty-ninth row and column
of \fig{fig:lcao_grid_c},
and the fourth and fifteenth rows and columns 
of \fig{fig:lcao_grid_n},
respectively.
For the \ion{C}{I} state, the rate coefficients are small
($\ratecoeff<10^{-13}\,\mathrm{cm^{3}\,s^{-1}}$),
owing to 
the avoided ionic crossing occurring at large internuclear distance
(around $80\,\abohr$, as mentioned above).
Concerning the \ion{N}{I} states, however, 
there are larger rates,
between the ground state and the 
\ion{N}{I} $\mathrm{2s^{2}.2p^{2}.3s\,^{4}P}$~state (row 4 -- column 5);
as well as between the first excited state
and the \ion{N}{I} 
$\mathrm{2s^{2}.2p^{2}.3p\,^{4}S^{o}\text{/}^{2}D^{o}\text{/}^{2}P^{o}}$~states
(row 15 -- columns 11/12/13).
In the context of non-LTE stellar spectroscopy, these
collisional processes are potentially important: 
the \ion{N}{I} $\mathrm{3s}$~and $\mathrm{3p}$~states
give rise to a number of optical and infra-red
transitions (see for example Table 2 of \citealt{1990A&amp;A...232..225G});
then, in so far as \ion{N}{I}~lines drive the departures from LTE,
these collisional processes would have a direct impact on 
the overall statistical equilibrium.

The \ion{C}{I} 
$\mathrm{2s.2p^{3}\,^{5}S^{o}\text{/}^{3}D^{o}\text{/}^{3}P^{o}}$~states, 
and the \ion{N}{I} $\mathrm{2s.2p^{4}\,^{3}D}$~state, enter into the model
only via two-electron processes involving ground-state hydrogen.
They correspond to the fourth, seventh, and fourteenth rows and columns
of \fig{fig:lcao_grid_c},
and the seventh row and column
of \fig{fig:lcao_grid_n},
respectively.
The rate coefficients involving the
\ion{C}{I} $\mathrm{2s.2p^{3}\,^{5}S^{o}}$~state are all zero,
because no avoided ionic crossings are predicted at these
internuclear distances, as mentioned above.
For the other states, the figures show
that the transitions are generally non-zero. 
Their rate coefficients are generally smaller
(by around a factor of ten) than those of single-electron transitions 
of similar excitation potentials and transition energies.
Nevertheless they can still
reach moderately high values
($\ratecoeff>10^{-10}\,\mathrm{cm^{3}\,s^{-1}}$).
Thus, in the context of non-LTE stellar spectroscopy, these
collisional processes are also potentially important.
However, as we discussed in \sect{method_lcao}, our current implementation
is not ideal, with rough approximations
made for the two-electron coupling probability, and the 
angular momentum couplings,
binding energies, and coefficients of fractional parentage 
of the covalent states.
If these transitions are confirmed to be influential
on the statistical equilibrium in stellar atmospheres, 
these calculations should be revisited in a future work.

Lastly, the (mostly blank) fourteenth row and column 
in \fig{fig:lcao_grid_n}
corresponds to the \ion{N}{I} $\mathrm{2s^{2}.2p^{2}.3s\,^{2}D}$~state.
This state couples to the excited ionic core 
$\mathrm{N^{+}\,2p^{2}\,^{1}D}$, and the current model predicts
an efficient charge transfer process.
This state is also coupled to the low-energy
\ion{N}{I} $\mathrm{2s^{2}.2p^{3}\,^{2}D^{o}}$~state, 
albeit inefficiently, via a two-electron
process involving excited hydrogen.

%-------------------------------------------------------------------------------
\section{Conclusion}
\label{conclusion}

We have presented new calculations for inelastic collisions with neutral
hydrogen, for \ion{C}{I} and \ion{N}{I},
using the LCAO model of \citet{2016PhRvA..93d2705B}.
This is based on more realistic physics than the approaches
commonly used for these species in the current non-LTE literature,
including for example the recipe of 
\citet{1968ZPhy..211..404D,1969ZPhy..225..483D}.
The rate coefficients for excitation and charge transfer
can be found as online tables at the CDS;
they are also made available at 
\url{https://github.com/barklem/public-data}.

Based on earlier work where comparisons could be done with full quantum
calculations, as well as indications from calculations using alternate
couplings \citep{2016PhRvA..93d2705B},
the uncertainties for the largest rates from these LCAO
calculations are expected to be about one order of magnitude, with
uncertainties becoming progressively larger as the rates become smaller.
As discussed above,
transitions involving two-electron processes will have larger uncertainties
than those involving
one-electron processes, and the results presented
here based on two-electron processes should 
at present only be considered indicative;
these processes should be studied more carefully in a future work.

We caution that there is some empirical evidence that 
the avoided ionic crossing mechanism considered in this work,
may not always be the dominant one, and thus that the LCAO results
presented here may underestimate the total
rate coefficients, for some transitions.
In earlier work, using the LCAO results alone that are
presented in this work and in \cite{2018A&A...610A..57B},
we found that the centre-to-limb variations
of lines of \ion{C}{I} \citep{2019A&A...624A.111A} and
\ion{O}{I} \citep{2018A&A...616A..89A},
precisely observed across the Sun,
could not be reproduced by our best models of the solar spectrum.

While this failure could be due to deficiencies in our adopted
three-dimensional (3D) hydrodynamic model solar 
atmosphere or 3D non-LTE radiative transfer,
a possible explanation is that the total O+H and C+H inelastic
rate coefficients were being underestimated.
In order to reproduce the observations, 
the LCAO rate coefficients were supplemented, using the
the free electron formulation of
\citet{1985JPhB...18L.167K,kaulakys1986free,1991JPhB...24L.127K}.
This method considers an alternative mechanism,
whereby inelastic processes take place via direct energy and
momentum transfer of the (free) active electron on the perturber
(as opposed to via electron transfer
in the LCAO model). A caveat is that this free electron
formulation is strictly valid only in the Rydberg-limit.
In summary, further research is still needed into the possible importance
of additional mechanisms, including alternative ionic channels.

%-------------------------------------------------------------------------------
\begin{acknowledgements}
The authors wish to thank the anonymous referee for 
helpful feedback on the original manuscript.
AMA acknowledges funds from the Alexander von Humboldt Foundation in the
framework of the Sofja Kovalevskaja Award endowed by the Federal Ministry of
Education and Research.
PSB acknowledges support from the Swedish Research Council and the 
project grant ``The New Milky Way'' from the Knut and 
Alice Wallenberg Foundation.
\end{acknowledgements}
%-------------------------------------------------------------------------------
\bibliographystyle{aa} 
\bibliography{/Users/ama51/Documents/work/papers/allpapers/bibl.bib}
%-------------------------------------------------------------------------------
\begin{appendix}
\section{Input data and symmetries considered}
\label{appendix_lcao}

\input{C_inp.tex}

\input{N_inp.tex}

\input{C_sym.tex}

\input{N_sym.tex}

\end{appendix}

%-------------------------------------------------------------------------------
\label{lastpage}
\end{document}

%% file: C_inp.tex
\begin{table*}
\begin{center}
\caption{Input data for the C+H calculations, sorted by the energies of the covalent CH states, $E_{\text{cov.}}$. Unless indicated, ground state hydrogen, $\mathrm{H(n=1)}$, is implied. Note that for the $\mathrm{2p^{3}\,^{5}S^{*}\text{/}^{3}D^{*}\text{/}^{3}P^{*}}$~states the binding energy of the valence electron, $E_{\text{lim.}}$, does not correspond to the energy of the target core; these states are included in the model via two-electron processes involving ground state hydrogen, as discussed in the text.}
\label{tab:C_inp}
\begin{tabular}{l l c c c c c c c c c c c} 
\hline
Index & 
State ($\mathrm{A+H}$) & 
$S_{\mathrm{A}}$ & 
$L_{\mathrm{A}}$ & 
$P_{\mathrm{A}}$ & 
$n_{\mathrm{A}}$ & 
$l_{\mathrm{A}}$ & 
$N_{\mathrm{A}}$ & 
$E_{\mathrm{A}} / eV$ & 
$E_{\text{cov.}} / eV$ & 
$E_{\text{lim.}} / eV$ & 
Core ($\mathrm{C}$) & 
$|G_{S_{\mathrm{A}},L_{\mathrm{A}}}^{S_{\mathrm{C}},L_{\mathrm{C}}}|$ \\
\hline
\hline
$           1$ & $\mathrm{2p^{2}\,^{3}P}$ & 1.0 & 1 & 0 & 2 & 1 & 2 &  0.003667 &  0.003667 & 11.265805 & $\mathrm{C^{+}\,2p\,^{2}P^{*}}$ &  1.0000 \\
$           2$ & $\mathrm{2p^{2}\,^{1}D}$ & 0.0 & 2 & 0 & 2 & 1 & 2 &  1.263725 &  1.263725 & 11.265805 & $\mathrm{C^{+}\,2p\,^{2}P^{*}}$ &  1.0000 \\
$           3$ & $\mathrm{2p^{2}\,^{1}S}$ & 0.0 & 0 & 0 & 2 & 1 & 2 &  2.684011 &  2.684011 & 11.265805 & $\mathrm{C^{+}\,2p\,^{2}P^{*}}$ &  1.0000 \\
$           4$ & $\mathrm{2p^{3}\,^{5}S^{o}}$ & 2.0 & 0 & 1 & 2 & 1 & 3 &  4.182632 &  4.182632 & 16.596210 & $\mathrm{C^{+}\,2p\,^{2}P^{*}}$ &  1.0000 \\
$           5$ & $\mathrm{3s\,^{3}P^{o}}$ & 1.0 & 1 & 1 & 3 & 0 & 1 &  7.485298 &  7.485298 & 11.265805 & $\mathrm{C^{+}\,2p\,^{2}P^{*}}$ &  1.0000 \\
$           6$ & $\mathrm{3s\,^{1}P^{o}}$ & 0.0 & 1 & 1 & 3 & 0 & 1 &  7.684766 &  7.684766 & 11.265805 & $\mathrm{C^{+}\,2p\,^{2}P^{*}}$ &  1.0000 \\
$           7$ & $\mathrm{2p^{3}\,^{3}D^{o}}$ & 1.0 & 2 & 1 & 2 & 1 & 3 &  7.946004 &  7.946004 & 20.550571 & $\mathrm{C^{+}\,2p\,^{2}P^{*}}$ &  1.0000 \\
$           8$ & $\mathrm{3p\,^{1}P}$ & 0.0 & 1 & 0 & 3 & 1 & 1 &  8.537097 &  8.537097 & 11.265805 & $\mathrm{C^{+}\,2p\,^{2}P^{*}}$ &  1.0000 \\
$           9$ & $\mathrm{3p\,^{3}D}$ & 1.0 & 2 & 0 & 3 & 1 & 1 &  8.644426 &  8.644426 & 11.265805 & $\mathrm{C^{+}\,2p\,^{2}P^{*}}$ &  1.0000 \\
$          10$ & $\mathrm{3p\,^{3}S}$ & 1.0 & 0 & 0 & 3 & 1 & 1 &  8.771132 &  8.771132 & 11.265805 & $\mathrm{C^{+}\,2p\,^{2}P^{*}}$ &  1.0000 \\
$          11$ & $\mathrm{3p\,^{3}P}$ & 1.0 & 1 & 0 & 3 & 1 & 1 &  8.849360 &  8.849360 & 11.265805 & $\mathrm{C^{+}\,2p\,^{2}P^{*}}$ &  1.0000 \\
$          12$ & $\mathrm{3p\,^{1}D}$ & 0.0 & 2 & 0 & 3 & 1 & 1 &  9.002582 &  9.002582 & 11.265805 & $\mathrm{C^{+}\,2p\,^{2}P^{*}}$ &  1.0000 \\
$          13$ & $\mathrm{3p\,^{1}S}$ & 0.0 & 0 & 0 & 3 & 1 & 1 &  9.171844 &  9.171844 & 11.265805 & $\mathrm{C^{+}\,2p\,^{2}P^{*}}$ &  1.0000 \\
$          14$ & $\mathrm{2p^{3}\,^{3}P^{o}}$ & 1.0 & 1 & 1 & 2 & 1 & 3 &  9.330422 &  9.330422 & 16.596210 & $\mathrm{C^{+}\,2p\,^{2}P^{*}}$ &  1.0000 \\
$          15$ & $\mathrm{3d\,^{1}D^{o}}$ & 0.0 & 2 & 1 & 3 & 2 & 1 &  9.631070 &  9.631070 & 11.265805 & $\mathrm{C^{+}\,2p\,^{2}P^{*}}$ &  1.0000 \\
$          16$ & $\mathrm{4s\,^{3}P^{o}}$ & 1.0 & 1 & 1 & 4 & 0 & 1 &  9.687238 &  9.687238 & 11.265805 & $\mathrm{C^{+}\,2p\,^{2}P^{*}}$ &  1.0000 \\
$          17$ & $\mathrm{3d\,^{3}F^{o}}$ & 1.0 & 3 & 1 & 3 & 2 & 1 &  9.698831 &  9.698831 & 11.265805 & $\mathrm{C^{+}\,2p\,^{2}P^{*}}$ &  1.0000 \\
$          18$ & $\mathrm{3d\,^{3}D^{o}}$ & 1.0 & 2 & 1 & 3 & 2 & 1 &  9.709173 &  9.709173 & 11.265805 & $\mathrm{C^{+}\,2p\,^{2}P^{*}}$ &  1.0000 \\
$          19$ & $\mathrm{4s\,^{1}P^{o}}$ & 0.0 & 1 & 1 & 4 & 0 & 1 &  9.712957 &  9.712957 & 11.265805 & $\mathrm{C^{+}\,2p\,^{2}P^{*}}$ &  1.0000 \\
$          20$ & $\mathrm{3d\,^{1}F^{o}}$ & 0.0 & 3 & 1 & 3 & 2 & 1 &  9.736432 &  9.736432 & 11.265805 & $\mathrm{C^{+}\,2p\,^{2}P^{*}}$ &  1.0000 \\
$          21$ & $\mathrm{3d\,^{1}P^{o}}$ & 0.0 & 1 & 1 & 3 & 2 & 1 &  9.761433 &  9.761433 & 11.265805 & $\mathrm{C^{+}\,2p\,^{2}P^{*}}$ &  1.0000 \\
$          22$ & $\mathrm{3d\,^{3}P^{o}}$ & 1.0 & 1 & 1 & 3 & 2 & 1 &  9.833789 &  9.833789 & 11.265805 & $\mathrm{C^{+}\,2p\,^{2}P^{*}}$ &  1.0000 \\
$          23$ & $\mathrm{4p\,^{1}P}$ & 0.0 & 1 & 0 & 4 & 1 & 1 &  9.988520 &  9.988520 & 11.265805 & $\mathrm{C^{+}\,2p\,^{2}P^{*}}$ &  1.0000 \\
$          24$ & $\mathrm{4p\,^{3}D}$ & 1.0 & 2 & 0 & 4 & 1 & 1 & 10.019545 & 10.019545 & 11.265805 & $\mathrm{C^{+}\,2p\,^{2}P^{*}}$ &  1.0000 \\
$          25$ & $\mathrm{4p\,^{3}S}$ & 1.0 & 0 & 0 & 4 & 1 & 1 & 10.055742 & 10.055742 & 11.265805 & $\mathrm{C^{+}\,2p\,^{2}P^{*}}$ &  1.0000 \\
$          26$ & $\mathrm{4p\,^{3}P}$ & 1.0 & 1 & 0 & 4 & 1 & 1 & 10.084161 & 10.084161 & 11.265805 & $\mathrm{C^{+}\,2p\,^{2}P^{*}}$ &  1.0000 \\
$          27$ & $\mathrm{4p\,^{1}D}$ & 0.0 & 2 & 0 & 4 & 1 & 1 & 10.138162 & 10.138162 & 11.265805 & $\mathrm{C^{+}\,2p\,^{2}P^{*}}$ &  1.0000 \\
$          28$ & $\mathrm{4p\,^{1}S}$ & 0.0 & 0 & 0 & 4 & 1 & 1 & 10.197912 & 10.197912 & 11.265805 & $\mathrm{C^{+}\,2p\,^{2}P^{*}}$ &  1.0000 \\
$          29$ & $\mathrm{2p^{2}\,^{3}P+H(n=2)}$ & 1.0 & 1 & 0 & 2 & 1 & 2 &  0.003667 & 10.202497 & 11.265805 & $\mathrm{C^{+}\,2p\,^{2}P^{*}}$ &  1.0000 \\
$          30$ & $\mathrm{4d\,^{1}D^{o}}$ & 0.0 & 2 & 1 & 4 & 2 & 1 & 10.352385 & 10.352385 & 11.265805 & $\mathrm{C^{+}\,2p\,^{2}P^{*}}$ &  1.0000 \\
$          31$ & $\mathrm{5s\,^{3}P^{o}}$ & 1.0 & 1 & 1 & 5 & 0 & 1 & 10.386466 & 10.386466 & 11.265805 & $\mathrm{C^{+}\,2p\,^{2}P^{*}}$ &  1.0000 \\
$          32$ & $\mathrm{4d\,^{3}F^{o}}$ & 1.0 & 3 & 1 & 4 & 2 & 1 & 10.386646 & 10.386646 & 11.265805 & $\mathrm{C^{+}\,2p\,^{2}P^{*}}$ &  1.0000 \\
$          33$ & $\mathrm{4d\,^{3}D^{o}}$ & 1.0 & 2 & 1 & 4 & 2 & 1 & 10.394774 & 10.394774 & 11.265805 & $\mathrm{C^{+}\,2p\,^{2}P^{*}}$ &  1.0000 \\
$          34$ & $\mathrm{5s\,^{1}P^{o}}$ & 0.0 & 1 & 1 & 5 & 0 & 1 & 10.404229 & 10.404229 & 11.265805 & $\mathrm{C^{+}\,2p\,^{2}P^{*}}$ &  1.0000 \\
$          35$ & $\mathrm{4d\,^{1}F^{o}}$ & 0.0 & 3 & 1 & 4 & 2 & 1 & 10.413529 & 10.413529 & 11.265805 & $\mathrm{C^{+}\,2p\,^{2}P^{*}}$ &  1.0000 \\
$          36$ & $\mathrm{4d\,^{1}P^{o}}$ & 0.0 & 1 & 1 & 4 & 2 & 1 & 10.418988 & 10.418988 & 11.265805 & $\mathrm{C^{+}\,2p\,^{2}P^{*}}$ &  1.0000 \\
$          37$ & $\mathrm{4d\,^{3}P^{o}}$ & 1.0 & 1 & 1 & 4 & 2 & 1 & 10.428242 & 10.428242 & 11.265805 & $\mathrm{C^{+}\,2p\,^{2}P^{*}}$ &  1.0000 \\
\hline
\multicolumn{13}{c}{Ionic states} \\
$          38$ & $\mathrm{2p\,^{2}P^{o}}$ & 0.5 & 1 & 1 & 2 & 1 & 1 & 11.265805 & - & - & - & - \\
\hline
\hline
\end{tabular}
\end{center}
\end{table*}

%% file: N_inp.tex
\begin{table*}
\begin{center}
\caption{Input data for the N+H calculations, sorted by the energies of the covalent NH states, $E_{\text{cov.}}$. Unless indicated, ground state hydrogen, $\mathrm{H(n=1)}$, is implied. Note that for the $\mathrm{2s.2p^{4}\,^{3}D}$~state the binding energy of the valence electron, $E_{\text{lim.}}$, does not correspond to the energy of the target core; these states are included in the model via two-electron processes involving ground state hydrogen, as discussed in the text.}
\label{tab:N_inp}
\begin{tabular}{l l c c c c c c c c c c c} 
\hline
Index & 
State ($\mathrm{A+H}$) & 
$S_{\mathrm{A}}$ & 
$L_{\mathrm{A}}$ & 
$P_{\mathrm{A}}$ & 
$n_{\mathrm{A}}$ & 
$l_{\mathrm{A}}$ & 
$N_{\mathrm{A}}$ & 
$E_{\mathrm{A}} / eV$ & 
$E_{\text{cov.}} / eV$ & 
$E_{\text{lim.}} / eV$ & 
Core ($\mathrm{C}$) & 
$|G_{S_{\mathrm{A}},L_{\mathrm{A}}}^{S_{\mathrm{C}},L_{\mathrm{C}}}|$ \\
\hline
\hline
$           1$ & $\mathrm{2p^{3}\,^{4}S^{o}}$ & 1.5 & 0 & 1 & 2 & 1 & 3 &  0.000000 &  0.000000 & 14.545154 & $\mathrm{N^{+}\,2p^{2}\,^{3}P}$ &  1.0000 \\
$           2$ & $\mathrm{2p^{3}\,^{2}D^{o}}$ & 0.5 & 2 & 1 & 2 & 1 & 3 &  2.383950 &  2.383950 & 14.545154 & $\mathrm{N^{+}\,2p^{2}\,^{3}P}$ &  0.7071 \\
$           3$ & $\mathrm{2p^{3}\,^{2}D^{o}}$ & 0.5 & 2 & 1 & 2 & 1 & 3 &  2.383950 &  2.383950 & 16.433100 & $\mathrm{N^{+}\,2p^{2}\,^{1}D}$ &  0.7071 \\
$           4$ & $\mathrm{2p^{3}\,^{2}P^{o}}$ & 0.5 & 1 & 1 & 2 & 1 & 3 &  3.575604 &  3.575604 & 14.545154 & $\mathrm{N^{+}\,2p^{2}\,^{3}P}$ &  0.7071 \\
$           5$ & $\mathrm{2p^{3}\,^{2}P^{o}}$ & 0.5 & 1 & 1 & 2 & 1 & 3 &  3.575604 &  3.575604 & 16.433100 & $\mathrm{N^{+}\,2p^{2}\,^{1}D}$ &  0.5270 \\
$           6$ & $\mathrm{2p^{3}\,^{4}S^{o}+H(n=2)}$ & 1.5 & 0 & 1 & 2 & 1 & 3 &  0.000000 & 10.198829 & 14.545154 & $\mathrm{N^{+}\,2p^{2}\,^{3}P}$ &  1.0000 \\
$           7$ & $\mathrm{3s\,^{4}P}$ & 1.5 & 1 & 0 & 3 & 0 & 1 & 10.332535 & 10.332535 & 14.545154 & $\mathrm{N^{+}\,2p^{2}\,^{3}P}$ &  1.0000 \\
$           8$ & $\mathrm{3s\,^{2}P}$ & 0.5 & 1 & 0 & 3 & 0 & 1 & 10.686888 & 10.686888 & 14.545154 & $\mathrm{N^{+}\,2p^{2}\,^{3}P}$ &  1.0000 \\
$           9$ & $\mathrm{2p^{4}\,^{4}P}$ & 1.5 & 1 & 0 & 2 & 1 & 4 & 10.926844 & 10.926844 & 25.970992 & $\mathrm{N^{+}\,2p^{2}\,^{3}P}$ &  1.0000 \\
$          10$ & $\mathrm{2p^{4}\,^{4}P}$ & 1.5 & 1 & 0 & 2 & 1 & 4 & 10.926844 & 10.926844 & 25.970992 & $\mathrm{N^{+}\,2p^{2}\,^{1}D}$ &  1.0000 \\
$          11$ & $\mathrm{3p\,^{2}S^{o}}$ & 0.5 & 0 & 1 & 3 & 1 & 1 & 11.602633 & 11.602633 & 14.545154 & $\mathrm{N^{+}\,2p^{2}\,^{3}P}$ &  1.0000 \\
$          12$ & $\mathrm{3p\,^{4}D^{o}}$ & 1.5 & 2 & 1 & 3 & 1 & 1 & 11.758641 & 11.758641 & 14.545154 & $\mathrm{N^{+}\,2p^{2}\,^{3}P}$ &  1.0000 \\
$          13$ & $\mathrm{3p\,^{4}P^{o}}$ & 1.5 & 1 & 1 & 3 & 1 & 1 & 11.841881 & 11.841881 & 14.545154 & $\mathrm{N^{+}\,2p^{2}\,^{3}P}$ &  1.0000 \\
$          14$ & $\mathrm{3p\,^{4}S^{o}}$ & 1.5 & 0 & 1 & 3 & 1 & 1 & 11.995575 & 11.995575 & 14.545154 & $\mathrm{N^{+}\,2p^{2}\,^{3}P}$ &  1.0000 \\
$          15$ & $\mathrm{3p\,^{2}D^{o}}$ & 0.5 & 2 & 1 & 3 & 1 & 1 & 12.005929 & 12.005929 & 14.545154 & $\mathrm{N^{+}\,2p^{2}\,^{3}P}$ &  1.0000 \\
$          16$ & $\mathrm{3p\,^{2}P^{o}}$ & 0.5 & 1 & 1 & 3 & 1 & 1 & 12.125052 & 12.125052 & 14.545154 & $\mathrm{N^{+}\,2p^{2}\,^{3}P}$ &  1.0000 \\
$          17$ & $\mathrm{3s\,^{2}D}$ & 0.5 & 2 & 0 & 3 & 0 & 1 & 12.356713 & 12.356713 & 16.433100 & $\mathrm{N^{+}\,2p^{2}\,^{1}D}$ &  1.0000 \\
$          18$ & $\mathrm{2p^{3}\,^{2}D^{o}+H(n=2)}$ & 0.5 & 2 & 1 & 2 & 1 & 3 &  2.383950 & 12.582779 & 14.545154 & $\mathrm{N^{+}\,2p^{2}\,^{3}P}$ &  0.7071 \\
$          19$ & $\mathrm{2p^{3}\,^{2}D^{o}+H(n=2)}$ & 0.5 & 2 & 1 & 2 & 1 & 3 &  2.383950 & 12.582779 & 16.433100 & $\mathrm{N^{+}\,2p^{2}\,^{1}D}$ &  0.7071 \\
$          20$ & $\mathrm{4s\,^{4}P}$ & 1.5 & 1 & 0 & 4 & 0 & 1 & 12.856736 & 12.856736 & 14.545154 & $\mathrm{N^{+}\,2p^{2}\,^{3}P}$ &  1.0000 \\
$          21$ & $\mathrm{4s\,^{2}P}$ & 0.5 & 1 & 0 & 4 & 0 & 1 & 12.918978 & 12.918978 & 14.545154 & $\mathrm{N^{+}\,2p^{2}\,^{3}P}$ &  1.0000 \\
$          22$ & $\mathrm{3d\,^{2}P}$ & 0.5 & 1 & 0 & 3 & 2 & 1 & 12.972099 & 12.972099 & 14.545154 & $\mathrm{N^{+}\,2p^{2}\,^{3}P}$ &  1.0000 \\
$          23$ & $\mathrm{3d\,^{4}F}$ & 1.5 & 3 & 0 & 3 & 2 & 1 & 12.983688 & 12.983688 & 14.545154 & $\mathrm{N^{+}\,2p^{2}\,^{3}P}$ &  1.0000 \\
$          24$ & $\mathrm{3d\,^{4}P}$ & 1.5 & 1 & 0 & 3 & 2 & 1 & 12.999168 & 12.999168 & 14.545154 & $\mathrm{N^{+}\,2p^{2}\,^{3}P}$ &  1.0000 \\
$          25$ & $\mathrm{3d\,^{2}F}$ & 0.5 & 3 & 0 & 3 & 2 & 1 & 12.999906 & 12.999906 & 14.545154 & $\mathrm{N^{+}\,2p^{2}\,^{3}P}$ &  1.0000 \\
$          26$ & $\mathrm{3d\,^{4}D}$ & 1.5 & 2 & 0 & 3 & 2 & 1 & 13.019322 & 13.019322 & 14.545154 & $\mathrm{N^{+}\,2p^{2}\,^{3}P}$ &  1.0000 \\
$          27$ & $\mathrm{3d\,^{2}D}$ & 0.5 & 2 & 0 & 3 & 2 & 1 & 13.035010 & 13.035010 & 14.545154 & $\mathrm{N^{+}\,2p^{2}\,^{3}P}$ &  1.0000 \\
$          28$ & $\mathrm{4p\,^{2}S^{o}}$ & 0.5 & 0 & 1 & 4 & 1 & 1 & 13.201565 & 13.201565 & 14.545154 & $\mathrm{N^{+}\,2p^{2}\,^{3}P}$ &  1.0000 \\
$          29$ & $\mathrm{4p\,^{4}D^{o}}$ & 1.5 & 2 & 1 & 4 & 1 & 1 & 13.244657 & 13.244657 & 14.545154 & $\mathrm{N^{+}\,2p^{2}\,^{3}P}$ &  1.0000 \\
$          30$ & $\mathrm{4p\,^{4}P^{o}}$ & 1.5 & 1 & 1 & 4 & 1 & 1 & 13.268207 & 13.268207 & 14.545154 & $\mathrm{N^{+}\,2p^{2}\,^{3}P}$ &  1.0000 \\
$          31$ & $\mathrm{4p\,^{2}D^{o}}$ & 0.5 & 2 & 1 & 4 & 1 & 1 & 13.294300 & 13.294300 & 14.545154 & $\mathrm{N^{+}\,2p^{2}\,^{3}P}$ &  1.0000 \\
$          32$ & $\mathrm{4p\,^{4}S^{o}}$ & 1.5 & 0 & 1 & 4 & 1 & 1 & 13.321559 & 13.321559 & 14.545154 & $\mathrm{N^{+}\,2p^{2}\,^{3}P}$ &  1.0000 \\
$          33$ & $\mathrm{4p\,^{2}P^{o}}$ & 0.5 & 1 & 1 & 4 & 1 & 1 & 13.342725 & 13.342725 & 14.545154 & $\mathrm{N^{+}\,2p^{2}\,^{3}P}$ &  1.0000 \\
$          34$ & $\mathrm{5s\,^{4}P}$ & 1.5 & 1 & 0 & 5 & 0 & 1 & 13.624243 & 13.624243 & 14.545154 & $\mathrm{N^{+}\,2p^{2}\,^{3}P}$ &  1.0000 \\
$          35$ & $\mathrm{5s\,^{2}P}$ & 0.5 & 1 & 0 & 5 & 0 & 1 & 13.648602 & 13.648602 & 14.545154 & $\mathrm{N^{+}\,2p^{2}\,^{3}P}$ &  1.0000 \\
\hline
\multicolumn{13}{c}{Ionic states} \\
$          36$ & $\mathrm{N^{+}\,2p^{2}\,^{3}P}$ & 1.0 & 1 & 0 & 2 & 1 & 2 & 14.545154 & - & - & - & - \\
$          37$ & $\mathrm{N^{+}\,2p^{2}\,^{1}D}$ & 0.0 & 2 & 0 & 2 & 1 & 2 & 16.433100 & - & - & - & - \\
\hline
\hline
\end{tabular}
\end{center}
\end{table*}

%% file: C_sym.tex
\begin{table*}
\begin{center}
\caption{Possible symmetries for the CH molecular states.}
\label{tab:C_sym}
\begin{tabular}{l l l l} 
\hline
Index & 
State & 
$g_{\text{total}}$ & 
Symmetries \\
\hline
\hline
$\mathrm{1}$  & $\mathrm{2p^{2}\,^{3}P}$  & $\mathrm{18}$  & $\mathrm{^{2}\Sigma^-,\        ^{2}\Pi,\   ^{4}\Sigma^-,\        ^{4}\Pi}$  \\
$\mathrm{2}$  & $\mathrm{2p^{2}\,^{1}D}$  & $\mathrm{10}$  & $\mathrm{^{2}\Sigma^+,\        ^{2}\Pi,\     ^{2}\Delta}$  \\
$\mathrm{3}$  & $\mathrm{2p^{2}\,^{1}S}$  & $\mathrm{2}$  & $\mathrm{^{2}\Sigma^+}$  \\
$\mathrm{4}$  & $\mathrm{2p^{3}\,^{5}S^{o}}$  & $\mathrm{10}$  & $\mathrm{^{4}\Sigma^-,\   ^{6}\Sigma^-}$  \\
$\mathrm{5}$  & $\mathrm{3s\,^{3}P^{o}}$  & $\mathrm{18}$  & $\mathrm{^{2}\Sigma^+,\        ^{2}\Pi,\   ^{4}\Sigma^+,\        ^{4}\Pi}$  \\
$\mathrm{6}$  & $\mathrm{3s\,^{1}P^{o}}$  & $\mathrm{6}$  & $\mathrm{^{2}\Sigma^+,\        ^{2}\Pi}$  \\
$\mathrm{7}$  & $\mathrm{2p^{3}\,^{3}D^{o}}$  & $\mathrm{30}$  & $\mathrm{^{2}\Sigma^-,\        ^{2}\Pi,\     ^{2}\Delta,\   ^{4}\Sigma^-,\        ^{4}\Pi,\     ^{4}\Delta}$  \\
$\mathrm{8}$  & $\mathrm{3p\,^{1}P}$  & $\mathrm{6}$  & $\mathrm{^{2}\Sigma^-,\        ^{2}\Pi}$  \\
$\mathrm{9}$  & $\mathrm{3p\,^{3}D}$  & $\mathrm{30}$  & $\mathrm{^{2}\Sigma^+,\        ^{2}\Pi,\     ^{2}\Delta,\   ^{4}\Sigma^+,\        ^{4}\Pi,\     ^{4}\Delta}$  \\
$\mathrm{10}$  & $\mathrm{3p\,^{3}S}$  & $\mathrm{6}$  & $\mathrm{^{2}\Sigma^+,\   ^{4}\Sigma^+}$  \\
$\mathrm{11}$  & $\mathrm{3p\,^{3}P}$  & $\mathrm{18}$  & $\mathrm{^{2}\Sigma^-,\        ^{2}\Pi,\   ^{4}\Sigma^-,\        ^{4}\Pi}$  \\
$\mathrm{12}$  & $\mathrm{3p\,^{1}D}$  & $\mathrm{10}$  & $\mathrm{^{2}\Sigma^+,\        ^{2}\Pi,\     ^{2}\Delta}$  \\
$\mathrm{13}$  & $\mathrm{3p\,^{1}S}$  & $\mathrm{2}$  & $\mathrm{^{2}\Sigma^+}$  \\
$\mathrm{14}$  & $\mathrm{2p^{3}\,^{3}P^{o}}$  & $\mathrm{18}$  & $\mathrm{^{2}\Sigma^+,\        ^{2}\Pi,\   ^{4}\Sigma^+,\        ^{4}\Pi}$  \\
$\mathrm{15}$  & $\mathrm{3d\,^{1}D^{o}}$  & $\mathrm{10}$  & $\mathrm{^{2}\Sigma^-,\        ^{2}\Pi,\     ^{2}\Delta}$  \\
$\mathrm{16}$  & $\mathrm{4s\,^{3}P^{o}}$  & $\mathrm{18}$  & $\mathrm{^{2}\Sigma^+,\        ^{2}\Pi,\   ^{4}\Sigma^+,\        ^{4}\Pi}$  \\
$\mathrm{17}$  & $\mathrm{3d\,^{3}F^{o}}$  & $\mathrm{42}$  & $\mathrm{^{2}\Sigma^+,\        ^{2}\Pi,\     ^{2}\Delta,\       ^{2}\Phi,\   ^{4}\Sigma^+,\        ^{4}\Pi,\     ^{4}\Delta,\       ^{4}\Phi}$  \\
$\mathrm{18}$  & $\mathrm{3d\,^{3}D^{o}}$  & $\mathrm{30}$  & $\mathrm{^{2}\Sigma^-,\        ^{2}\Pi,\     ^{2}\Delta,\   ^{4}\Sigma^-,\        ^{4}\Pi,\     ^{4}\Delta}$  \\
$\mathrm{19}$  & $\mathrm{4s\,^{1}P^{o}}$  & $\mathrm{6}$  & $\mathrm{^{2}\Sigma^+,\        ^{2}\Pi}$  \\
$\mathrm{20}$  & $\mathrm{3d\,^{1}F^{o}}$  & $\mathrm{14}$  & $\mathrm{^{2}\Sigma^+,\        ^{2}\Pi,\     ^{2}\Delta,\       ^{2}\Phi}$  \\
$\mathrm{21}$  & $\mathrm{3d\,^{1}P^{o}}$  & $\mathrm{6}$  & $\mathrm{^{2}\Sigma^+,\        ^{2}\Pi}$  \\
$\mathrm{22}$  & $\mathrm{3d\,^{3}P^{o}}$  & $\mathrm{18}$  & $\mathrm{^{2}\Sigma^+,\        ^{2}\Pi,\   ^{4}\Sigma^+,\        ^{4}\Pi}$  \\
$\mathrm{23}$  & $\mathrm{4p\,^{1}P}$  & $\mathrm{6}$  & $\mathrm{^{2}\Sigma^-,\        ^{2}\Pi}$  \\
$\mathrm{24}$  & $\mathrm{4p\,^{3}D}$  & $\mathrm{30}$  & $\mathrm{^{2}\Sigma^+,\        ^{2}\Pi,\     ^{2}\Delta,\   ^{4}\Sigma^+,\        ^{4}\Pi,\     ^{4}\Delta}$  \\
$\mathrm{25}$  & $\mathrm{4p\,^{3}S}$  & $\mathrm{6}$  & $\mathrm{^{2}\Sigma^+,\   ^{4}\Sigma^+}$  \\
$\mathrm{26}$  & $\mathrm{4p\,^{3}P}$  & $\mathrm{18}$  & $\mathrm{^{2}\Sigma^-,\        ^{2}\Pi,\   ^{4}\Sigma^-,\        ^{4}\Pi}$  \\
$\mathrm{27}$  & $\mathrm{4p\,^{1}D}$  & $\mathrm{10}$  & $\mathrm{^{2}\Sigma^+,\        ^{2}\Pi,\     ^{2}\Delta}$  \\
$\mathrm{28}$  & $\mathrm{4p\,^{1}S}$  & $\mathrm{2}$  & $\mathrm{^{2}\Sigma^+}$  \\
$\mathrm{29}$  & $\mathrm{2p^{2}\,^{3}P+H(n=2)}$  & $\mathrm{72}$  & $\mathrm{^{2}\Sigma^-,\        ^{2}\Pi,\   ^{4}\Sigma^-,\        ^{4}\Pi,\   ^{2}\Sigma^+,\     ^{2}\Delta,\   ^{4}\Sigma^+,\     ^{4}\Delta}$  \\
$\mathrm{30}$  & $\mathrm{4d\,^{1}D^{o}}$  & $\mathrm{10}$  & $\mathrm{^{2}\Sigma^-,\        ^{2}\Pi,\     ^{2}\Delta}$  \\
$\mathrm{31}$  & $\mathrm{5s\,^{3}P^{o}}$  & $\mathrm{18}$  & $\mathrm{^{2}\Sigma^+,\        ^{2}\Pi,\   ^{4}\Sigma^+,\        ^{4}\Pi}$  \\
$\mathrm{32}$  & $\mathrm{4d\,^{3}F^{o}}$  & $\mathrm{42}$  & $\mathrm{^{2}\Sigma^+,\        ^{2}\Pi,\     ^{2}\Delta,\       ^{2}\Phi,\   ^{4}\Sigma^+,\        ^{4}\Pi,\     ^{4}\Delta,\       ^{4}\Phi}$  \\
$\mathrm{33}$  & $\mathrm{4d\,^{3}D^{o}}$  & $\mathrm{30}$  & $\mathrm{^{2}\Sigma^-,\        ^{2}\Pi,\     ^{2}\Delta,\   ^{4}\Sigma^-,\        ^{4}\Pi,\     ^{4}\Delta}$  \\
$\mathrm{34}$  & $\mathrm{5s\,^{1}P^{o}}$  & $\mathrm{6}$  & $\mathrm{^{2}\Sigma^+,\        ^{2}\Pi}$  \\
$\mathrm{35}$  & $\mathrm{4d\,^{1}F^{o}}$  & $\mathrm{14}$  & $\mathrm{^{2}\Sigma^+,\        ^{2}\Pi,\     ^{2}\Delta,\       ^{2}\Phi}$  \\
$\mathrm{36}$  & $\mathrm{4d\,^{1}P^{o}}$  & $\mathrm{6}$  & $\mathrm{^{2}\Sigma^+,\        ^{2}\Pi}$  \\
$\mathrm{37}$  & $\mathrm{4d\,^{3}P^{o}}$  & $\mathrm{18}$  & $\mathrm{^{2}\Sigma^+,\        ^{2}\Pi,\   ^{4}\Sigma^+,\        ^{4}\Pi}$  \\
$\mathrm{38}$  & $\mathrm{C^{+}\,2p\,^{2}P^{o}}$  & $\mathrm{6}$  & $\mathrm{^{2}\Sigma^+,\        ^{2}\Pi}$  \\
&&&\\ 
     \multicolumn{3}{l}{Number of symmetries to calculate :   2} & $    ^{2}\Sigma^+,\        ^{2}\Pi$ \\
 \multicolumn{3}{l}{$g:$} & $              2,\              4$ \\
\hline
\hline
\end{tabular}
\end{center}
\end{table*}

%% file: N_sym.tex
\begin{table*}
\begin{center}
\caption{Possible symmetries for the NH molecular states.}
\label{tab:N_sym}
\begin{tabular}{l l l l} 
\hline
Index & 
State & 
$g_{\text{total}}$ & 
Symmetries \\
\hline
\hline
$\mathrm{1}$  & $\mathrm{2p^{3}\,^{4}S^{o}}$  & $\mathrm{8}$  & $\mathrm{^{3}\Sigma^-,\   ^{5}\Sigma^-}$  \\
$\mathrm{2}$  & $\mathrm{2p^{3}\,^{2}D^{o}}$  & $\mathrm{20}$  & $\mathrm{^{1}\Sigma^-,\        ^{1}\Pi,\     ^{1}\Delta,\   ^{3}\Sigma^-,\        ^{3}\Pi,\     ^{3}\Delta}$  \\
$\mathrm{3}$  & $\mathrm{2p^{3}\,^{2}P^{o}}$  & $\mathrm{12}$  & $\mathrm{^{1}\Sigma^+,\        ^{1}\Pi,\   ^{3}\Sigma^+,\        ^{3}\Pi}$  \\
$\mathrm{4}$  & $\mathrm{2p^{3}\,^{4}S^{o}+H(n=2)}$  & $\mathrm{32}$  & $\mathrm{^{3}\Sigma^-,\   ^{5}\Sigma^-,\        ^{3}\Pi,\        ^{5}\Pi}$  \\
$\mathrm{5}$  & $\mathrm{3s\,^{4}P}$  & $\mathrm{24}$  & $\mathrm{^{3}\Sigma^-,\        ^{3}\Pi,\   ^{5}\Sigma^-,\        ^{5}\Pi}$  \\
$\mathrm{6}$  & $\mathrm{3s\,^{2}P}$  & $\mathrm{12}$  & $\mathrm{^{1}\Sigma^-,\        ^{1}\Pi,\   ^{3}\Sigma^-,\        ^{3}\Pi}$  \\
$\mathrm{7}$  & $\mathrm{2p^{4}\,^{4}P}$  & $\mathrm{24}$  & $\mathrm{^{3}\Sigma^-,\        ^{3}\Pi,\   ^{5}\Sigma^-,\        ^{5}\Pi}$  \\
$\mathrm{8}$  & $\mathrm{3p\,^{2}S^{o}}$  & $\mathrm{4}$  & $\mathrm{^{1}\Sigma^-,\   ^{3}\Sigma^-}$  \\
$\mathrm{9}$  & $\mathrm{3p\,^{4}D^{o}}$  & $\mathrm{40}$  & $\mathrm{^{3}\Sigma^-,\        ^{3}\Pi,\     ^{3}\Delta,\   ^{5}\Sigma^-,\        ^{5}\Pi,\     ^{5}\Delta}$  \\
$\mathrm{10}$  & $\mathrm{3p\,^{4}P^{o}}$  & $\mathrm{24}$  & $\mathrm{^{3}\Sigma^+,\        ^{3}\Pi,\   ^{5}\Sigma^+,\        ^{5}\Pi}$  \\
$\mathrm{11}$  & $\mathrm{3p\,^{4}S^{o}}$  & $\mathrm{8}$  & $\mathrm{^{3}\Sigma^-,\   ^{5}\Sigma^-}$  \\
$\mathrm{12}$  & $\mathrm{3p\,^{2}D^{o}}$  & $\mathrm{20}$  & $\mathrm{^{1}\Sigma^-,\        ^{1}\Pi,\     ^{1}\Delta,\   ^{3}\Sigma^-,\        ^{3}\Pi,\     ^{3}\Delta}$  \\
$\mathrm{13}$  & $\mathrm{3p\,^{2}P^{o}}$  & $\mathrm{12}$  & $\mathrm{^{1}\Sigma^+,\        ^{1}\Pi,\   ^{3}\Sigma^+,\        ^{3}\Pi}$  \\
$\mathrm{14}$  & $\mathrm{3s\,^{2}D}$  & $\mathrm{20}$  & $\mathrm{^{1}\Sigma^+,\        ^{1}\Pi,\     ^{1}\Delta,\   ^{3}\Sigma^+,\        ^{3}\Pi,\     ^{3}\Delta}$  \\
$\mathrm{15}$  & $\mathrm{2p^{3}\,^{2}D^{o}+H(n=2)}$  & $\mathrm{80}$  & $\mathrm{^{1}\Sigma^-,\        ^{1}\Pi,\     ^{1}\Delta,\   ^{3}\Sigma^-,\        ^{3}\Pi,\     ^{3}\Delta,\   ^{1}\Sigma^+,\       ^{1}\Phi,\   ^{3}\Sigma^+,\       ^{3}\Phi}$  \\
$\mathrm{16}$  & $\mathrm{4s\,^{4}P}$  & $\mathrm{24}$  & $\mathrm{^{3}\Sigma^-,\        ^{3}\Pi,\   ^{5}\Sigma^-,\        ^{5}\Pi}$  \\
$\mathrm{17}$  & $\mathrm{4s\,^{2}P}$  & $\mathrm{12}$  & $\mathrm{^{1}\Sigma^-,\        ^{1}\Pi,\   ^{3}\Sigma^-,\        ^{3}\Pi}$  \\
$\mathrm{18}$  & $\mathrm{3d\,^{2}P}$  & $\mathrm{12}$  & $\mathrm{^{1}\Sigma^-,\        ^{1}\Pi,\   ^{3}\Sigma^-,\        ^{3}\Pi}$  \\
$\mathrm{19}$  & $\mathrm{3d\,^{4}F}$  & $\mathrm{56}$  & $\mathrm{^{3}\Sigma^-,\        ^{3}\Pi,\     ^{3}\Delta,\       ^{3}\Phi,\   ^{5}\Sigma^-,\        ^{5}\Pi,\     ^{5}\Delta,\       ^{5}\Phi}$  \\
$\mathrm{20}$  & $\mathrm{3d\,^{4}P}$  & $\mathrm{24}$  & $\mathrm{^{3}\Sigma^-,\        ^{3}\Pi,\   ^{5}\Sigma^-,\        ^{5}\Pi}$  \\
$\mathrm{21}$  & $\mathrm{3d\,^{2}F}$  & $\mathrm{28}$  & $\mathrm{^{1}\Sigma^-,\        ^{1}\Pi,\     ^{1}\Delta,\       ^{1}\Phi,\   ^{3}\Sigma^-,\        ^{3}\Pi,\     ^{3}\Delta,\       ^{3}\Phi}$  \\
$\mathrm{22}$  & $\mathrm{3d\,^{4}D}$  & $\mathrm{40}$  & $\mathrm{^{3}\Sigma^+,\        ^{3}\Pi,\     ^{3}\Delta,\   ^{5}\Sigma^+,\        ^{5}\Pi,\     ^{5}\Delta}$  \\
$\mathrm{23}$  & $\mathrm{3d\,^{2}D}$  & $\mathrm{20}$  & $\mathrm{^{1}\Sigma^+,\        ^{1}\Pi,\     ^{1}\Delta,\   ^{3}\Sigma^+,\        ^{3}\Pi,\     ^{3}\Delta}$  \\
$\mathrm{24}$  & $\mathrm{4p\,^{2}S^{o}}$  & $\mathrm{4}$  & $\mathrm{^{1}\Sigma^-,\   ^{3}\Sigma^-}$  \\
$\mathrm{25}$  & $\mathrm{4p\,^{4}D^{o}}$  & $\mathrm{40}$  & $\mathrm{^{3}\Sigma^-,\        ^{3}\Pi,\     ^{3}\Delta,\   ^{5}\Sigma^-,\        ^{5}\Pi,\     ^{5}\Delta}$  \\
$\mathrm{26}$  & $\mathrm{4p\,^{4}P^{o}}$  & $\mathrm{24}$  & $\mathrm{^{3}\Sigma^+,\        ^{3}\Pi,\   ^{5}\Sigma^+,\        ^{5}\Pi}$  \\
$\mathrm{27}$  & $\mathrm{4p\,^{2}D^{o}}$  & $\mathrm{20}$  & $\mathrm{^{1}\Sigma^-,\        ^{1}\Pi,\     ^{1}\Delta,\   ^{3}\Sigma^-,\        ^{3}\Pi,\     ^{3}\Delta}$  \\
$\mathrm{28}$  & $\mathrm{4p\,^{4}S^{o}}$  & $\mathrm{8}$  & $\mathrm{^{3}\Sigma^-,\   ^{5}\Sigma^-}$  \\
$\mathrm{29}$  & $\mathrm{4p\,^{2}P^{o}}$  & $\mathrm{12}$  & $\mathrm{^{1}\Sigma^+,\        ^{1}\Pi,\   ^{3}\Sigma^+,\        ^{3}\Pi}$  \\
$\mathrm{30}$  & $\mathrm{5s\,^{4}P}$  & $\mathrm{24}$  & $\mathrm{^{3}\Sigma^-,\        ^{3}\Pi,\   ^{5}\Sigma^-,\        ^{5}\Pi}$  \\
$\mathrm{31}$  & $\mathrm{5s\,^{2}P}$  & $\mathrm{12}$  & $\mathrm{^{1}\Sigma^-,\        ^{1}\Pi,\   ^{3}\Sigma^-,\        ^{3}\Pi}$  \\
$\mathrm{32}$  & $\mathrm{N^{+}\,2p^{2}\,^{3}P}$  & $\mathrm{9}$  & $\mathrm{^{3}\Sigma^-,\        ^{3}\Pi}$  \\
$\mathrm{33}$  & $\mathrm{N^{+}\,2p^{2}\,^{1}D}$  & $\mathrm{5}$  & $\mathrm{^{1}\Sigma^+,\        ^{1}\Pi,\     ^{1}\Delta}$  \\
&&&\\ 
     \multicolumn{3}{l}{Number of symmetries to calculate :   5} & $    ^{3}\Sigma^-,\        ^{3}\Pi,\   ^{1}\Sigma^+,\        ^{1}\Pi,\     ^{1}\Delta$ \\
 \multicolumn{3}{l}{$g:$} & $              3,\              6,\              1,\              2,\              2$ \\
\hline
\hline
\end{tabular}
\end{center}
\end{table*}